\documentclass[twocolumn,apl,epsf,superscriptaddress]{revtex4-2}

\usepackage{amsmath}
\usepackage{amsfonts}
\usepackage{amssymb}
\usepackage{epsfig}
\usepackage{epsf}
\usepackage{array}
\usepackage{color}
\usepackage{ulem}

\newcommand{\vect}[1]{\mathbf #1}
\newcommand{\vecs}[1]{\mbox{\boldmath $#1$}}
\newcommand{\e}{\mathrm{e}}
\renewcommand{\i}{\mathrm{i}}
\usepackage{setspace}

\renewcommand{\bibsection}{\section*{R\lowercase{eferences}}}

\begin{document}


%
\title{Critical enhancement of the spin Hall effect by spin fluctuations}
%
\author{Satoshi Okamoto}
\altaffiliation{okapon@ornl.gov}
\affiliation{Materials Science and Technology Division, Oak Ridge National Laboratory, Oak Ridge, Tennessee 37831, USA}
%
\author{Naoto Nagaosa}
\affiliation{RIKEN Center for Emergent Matter Science (CEMS), Wako, Saitama 351-0198, Japan}

\begin{abstract}
The spin Hall (SH) effect, the conversion of the electric current to the spin current along the transverse direction, relies on the relativistic spin-orbit coupling (SOC). 
Here, we develop a microscopic theory on the mechanisms of the SH effect in magnetic metals, where itinerant electrons are coupled with localized magnetic moments via the Hund exchange interaction and the SOC. 
Both antiferromagnetic metals and ferromagnetic metals are considered. 
It is shown that the SH conductivity can be significantly enhanced by the spin fluctuation when approaching the magnetic transition temperature of both cases. 
For antiferromagnetic metals, the pure SH effect appears in the entire temperature range, while for ferromagnetic metals, the pure SH effect is expected to be replaced by the anomalous Hall effect below the transition temperature.  
We discuss possible experimental realizations and the effect of the quantum criticality when the antiferromagnetic transition temperature is tuned to zero temperature.
\end{abstract}
\maketitle


The spin Hall (SH) effect \cite{Dyakonov1971a,Dyakonov1971b,Hirsch1999} and its reciprocal effect, the inverse SH effect \cite{Saitoh2006}, 
are among the most important components for the spintronic application \cite{Maekawa2006}  
because they allow the  
electrical conversion between charge current and pure spin current, where electrons with opposite spin components flow along opposite directions with zero net charge current. 
Based on these effects, a variety of phenomena have been envisioned \cite{Murakami2011,Sinova2015}. 
The SH effect and the anomalous Hall (AH) effect ~\cite{Nagaosa2010} are both rooted in the relativistic spin-orbit coupling (SOC), and
these effects are traditionally understood 
as arising from intrinsic mechanisms, i.e., band effects, \cite{Sinova2004,Murakami2004} or extrinsic mechanisms, i.e., impurity/disorder effects \cite{Smit1955,Smit1958,Berger1972,Berger1970,Crepieux2001,Tse2006} and interfaces \cite{Wang2016}. 

There have appeared a number of proposals of the extrinsic mechanisms of the SH effect utilizing excitations or fluctuations in solids, 
such as phonons \cite{Gorini2015,Karnad2018,Xiao2019}. 
Identifying new mechanisms thus opens up a new research avenue and hereby helps to improve the efficiency of the SH effect, 
which remains small for practical applications \cite{Hoffmann2013}. 
Recently, the current authors proposed extrinsic mechanisms focusing on the spin fluctuation (SF) 
in nearly ferromagnetic (FM) disordered systems \cite{Okamoto2019}. 
In these mechanisms, the critical SF associated with the zero-temperature FM quantum critical point (QCP) plays a fundamental role. 
It was predicted that the SH conductivity $\sigma_{\rm SH}$ is maximized at nonzero temperature when approaching the QCP. 
When the FM transition temperature $T_{\rm C}$ is finite, the pure SH effect is replaced by the AH effect below $T_{\rm C}$, 
thus limiting the operation temperature range of the SH effect. 
This limitation could be lifted when the antiferromagnetic (AFM) SF is considered because the net magnetic moment is absent 
even below the AFM transition temperature $T_{\rm N}$. 
From the study of itinerant electron magnetism \cite{Moriya1985}, it has been recognized that 
the FM SF and the AFM SF provide qualitatively different behavior in electronic specific heat, conductivity, {\it etc} \cite{Ueda1975,Ueda1977,Moriya2000}. 
Thus, the SH effect could be another example that highlights the difference between FM SF and AFM SF. 

While the SH effect due to the FM critical SF has not been experimentally examined, yet, 
Refs.~\cite{Wei2012,Ou2018,Wu2022} examined the SH effect in FM alloys with finite $T_{\rm C}$. 
In Ref.~\cite{Wei2012}, Wei et al. reported that the temperature dependence of the inverse SH resistance of Ni-Pd alloys follows the uniform second-order nonlinear susceptibility $\chi_2$, 
but the  inverse SH resistance has peaks above and below the Curie temperature and changes its sign at $T_{\rm C}$. 
This behavior is consistent with the theoretical prediction in Ref.~\cite{Gu2012}, which used a static mean field approximation to the model proposed by Kondo \cite{Kondo1962}. 
On the other hand in Ref.~\cite{Ou2018}, Ou et al. reported that the inverse SH effect of Fe-Pt alloys is maximized near $T_{\rm C}$ 
as if it follows the uniform linear susceptibility. 
More recently, Wu et al. reported similar effects using Ni-Cu alloys \cite{Wu2022}. 
For AFM systems, early work on Cr has already reported the large SH effects \cite{Du2014,Qu2015}. 
Recently, Fang and coworkers found that the SH conductivity in metallic Cr is enhanced when temperature is approaching the N{\'e}el temperature $T_{\rm N}$ \cite{Fang2023}, 
suggesting the AFM SF as the main mechanism of the SH effect. 
However, the effect of AFM SF to the SH effect has not been theoretically addressed.

The main purpose of this work is to develop the theoretical description of the SH effect in magnetic metallic systems by the SF when the magnetic transition temperature ($T_{\rm C}$ or $T_{\rm N}$) is finite. 
Our theory is based on a microscopic model 
describing the coupling between itinerant electrons and localized magnetic moments by Kondo \cite{Kondo1962} and 
the self-consistent renormalization theory describing the fluctuation of localized moments by Moriya \cite{Moriya1985}. 
The main difference between AFM systems and FM systems is that the AFM ordering or correlation is characterized by the nonzero magnetic wave vector $\vect Q$. 
Thus, itinerant electrons scattered by the AFM SF gain or lose corresponding momentum. 
Our theory takes into account this momentum conservation appropriately. 
Despite this difference, it is demonstrated that the SH conductivity is enhanced as temperature is approaching $T_{\rm C}$ for FM systems or $T_{\rm N}$ for AFM systems.  
The result for the AFM systems strongly supports the conjecture made in Ref.~\cite{Fang2023}. 
We highlight the qualitatively different behavior between the AFM SF and the FM SF 
near the finite-temperature phase transition and near the QCP. 

For magnetic metallic systems, intrinsic mechanisms could also contribute to a variety of Hall effects, such as
the AH effect by the Berry curvature and the topological Hall effect induced by chiral spin ordering. 
This work, however, does not cover these effects because these are band effects and do not show diverging behavior. 

\section*{R\lowercase{esults}}

In this section, we present our main results. 
The first subsection is devoted to setting up our theoretical tools. 
We introduce the $s$-$d$ Hamiltonian, which is modified from the original form developed by Kondo \cite{Kondo1962}. 
Based on this model, the scattering mechanism for the SH conductivity will be clarified. 
Then, we set up the Gaussian action, which describes the 
SF based on the self-consistent renormalization theory by Moriya \cite{Moriya1985}. 
In the second subsection, theoretical results of the SH conductivity will be presented. 

\begin{figure}
\begin{center}
\includegraphics[width=0.9\columnwidth, clip]{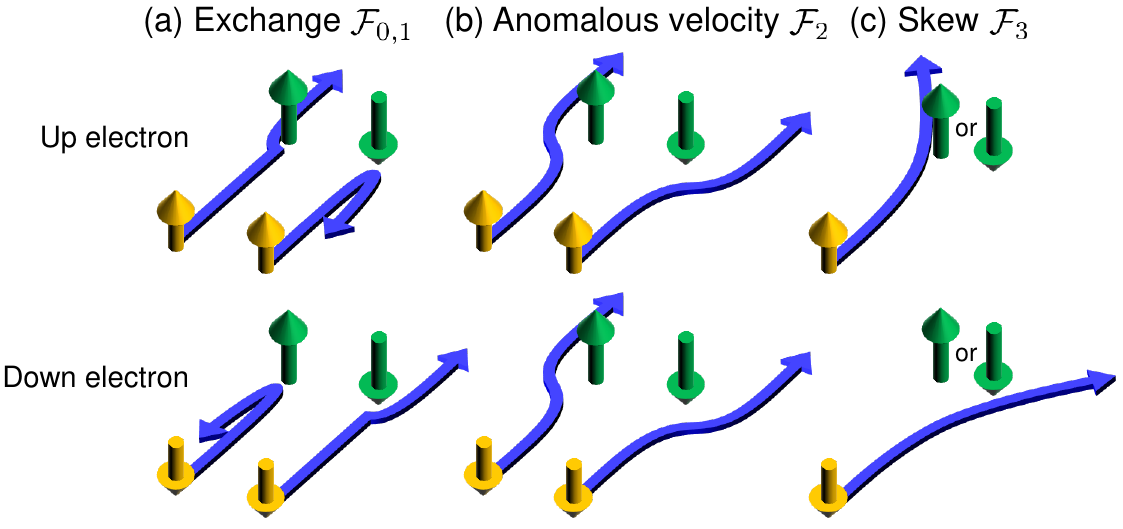}
\caption{Schematic view of the Kondo model.   
(a) Spin dependent scattering by ${\cal F}_{0,1}$ terms, (b) anomalous velocity induced by ${\cal F}_2$ terms, and (c) skew scattering by ${\cal F}_3$terms. 
Yellow arrows indicate conduction electrons, and green arrows indicate local moments. 
In the ${\cal F}_{2 (3)}$ scattering processes, the electron deflection depends on the direction of the local moment (the electron spin), 
eading to the side-jump (skew-scattering) contribution to $\sigma_{\rm SH}$. 
Adapted from Okamoto, S. Egami, T. \& Nagaosa, N. Phys. Rev. Lett. {\bf 123}, 196603 (2019).}
\label{fig:scatter}
\end{center}
\end{figure}

\subsection*{Theoretical model and formalism}

In this work, we consider three-dimensional systems. 
The $s$-$d$ Hamiltonian is written as $H = H_0 + H_{\rm K}$. 
Here, the non-interacting itinerant electron part is described by 
$H_0=\sum_{\vect k, \nu} \varepsilon_{\vect k} a_{\vect k \nu}^\dag a_{\vect k \nu}$, 
where
$\varepsilon_{\vect k}$ is an energy eigenvalue at momentum $\vect k$ given by 
$\varepsilon_{\vect k} = \frac{\hbar^2 k^2}{2m} - \varepsilon_{\rm F}$ with the Fermi energy $\varepsilon_{\rm F}$ and the carrier effective mass $m$, and 
$a_{\vect k \nu}^{(\dag)}$ is an annihilation (creation) operator of an electron at momentum $\vect k$ with the spin index $\nu$. 
We chose the simplest-possible band structure as it allows detailed analytical calculations. 
This electronic part of the Hamiltonian is assumed to be unrenormalized \cite{Moriya1985}. 
However, as discussed briefly later, going beyond this assumption is necessary especially near the non-zero critical temperature. 
The $s$-$d$ coupling term in the original model has a mixed representation of momentum of conduction electrons and 
real-space coordinate of localized magnetic moments (see Ref.~\onlinecite{Kondo1962} and Supplementary Note 1 for details). 
For magnetic metallic systems, where localized magnetic moments form a periodic lattice and conduction electrons hop through the same lattice sites, 
it is more convenient to express the model Hamiltonian entirely in the momentum space as 
\begin{eqnarray}
H_{\rm K} \!\!\!&=&\!\!\!
- \frac {1}{\sqrt{N}} \sum_{\vect k, \vect k'} \sum_{\nu, \nu'} 
a_{\vect k \nu}^\dag a_{\vect k' \nu'} 
\Biggl[ 2(\vect J_{\vect k - \vect k'} \cdot \vect s_{\nu \nu'}) \nonumber \\
&\times&\!\!\! \bigl\{ {\cal F}_0  
+ 2 {\cal F}_1(\vect k \cdot \vect k') \bigr\} 
+ \i {\cal F}_2 \vect J_{\vect k - \vect k'} \cdot (\vect k' \times \vect k)  \nonumber \\
&+&\!\!\! \frac{ \i}{\sqrt{N}} \sum_{\vect p}
{\cal F}_3  
\biggl\{( \vect J_{\vect p} \cdot \vect s_{\nu \nu'}) \bigl( \vect J_{\vect k - \vect k' - \vect p} \cdot (\vect k' \times \vect k) \bigr)  \nonumber \\
&+&\!\!\! \bigl( \vect J_{\vect p} \cdot (\vect k' \times \vect k) \bigr) (\vect J_{\vect k - \vect k' -\vect p} \cdot \vect s_{\nu \nu'}) \nonumber \\
&-&\!\!\! \frac{2}{3} (\vect J_{\vect p} \cdot \vect J_{\vect k - \vect k' -\vect p}) \bigl( \vect s_{\nu \nu'} \cdot (\vect k' \times \vect k) \bigr) \biggr\} \Biggr] . 
\label{eq:Kondo}
\end{eqnarray}
Here, 
$\vect s_{\nu \nu'}=\frac{1}{2} \vecs \sigma_{\nu \nu'}$ is the spin of a conduction electron with the Pauli matrices $\vecs \sigma$. 
$N$ is the total number of lattice sites.
$\vect J_{\vect p}$ is the Fourier transform of a local spin moment $\vect J_n$ at position $\vect R_n$ defined by 
$\vect J_{\vect p} = \frac{1}{\sqrt{N}} \sum_n \vect J_n \e^{- \i \vect p \cdot \vect R_n}$. 
Parameters ${\cal F}_l$ \cite{Okamoto2019} are related to $F_l$ defined in Ref.~\cite{Kondo1962}. 
${\cal F}_{0,1}$ terms correspond to the standard $s$-$d$ exchange interaction or Hund coupling as depicted in Fig.~\ref{fig:scatter} (a).  
Note that the ${\cal F}_{0,1}$ terms represent ferromagnetic coupling. 
With these leading terms, the behavior of the theoretical model is less exotic than 
that with antiferromagnetic coupling often used in the context of heavy fermions \cite{Anderson1961,Kondo1964}. 
The subleading  
${\cal F}_{2,3}$ terms represent the exchange of angular momentum between a conduction electron and a local moment. 
These terms are odd (linear or cubic)-order in $\vect J_n$ and $\vect s$ 
and induce the electron deflection depending on the direction of $\vect J_n$ or $\vect s$. 
More precisely, the ${\cal F}_2$ term and the ${\cal F}_3$ term generate 
the side-jump and the skew-scattering contributions to the SH conductivity, respectively, as depicted in Fig.~\ref{fig:scatter} (b) and (c). 
From Eq.~(\ref{eq:Kondo}) and the position operator $\vect r$, the velocity operator is obtained as
$\vect v = (\i/\hbar) [H_{\rm K},\vect r]$. 
The anomalous velocity, the main source of the side-jump contribution, arises from 
the ${\cal F}_{2}$ term (see Supplementary Note 1 for details).

From Eq.~(\ref{eq:Kondo}), one can notice the main difference from FM alloy systems \cite{Okamoto2019}. 
Since itinerant electrons and localized moments share the same lattice structure, 
their coupling does not have a phase factor such as $\e^{ \i \vect p \cdot (\vect R_n - \vect R_{n'})}$, where 
$\vect R_n$ is the position of the localized moment $\vect J_n$. 
Therefore, the SF could contribute to the SH effect even if 
it has characteristic momentum $\vect Q \ne  \vect 0$, such as in AFM systems, 
without introducing destructive effects. 
Otherwise, averaging over the lattice coordinate would lead to zero SH effect as 
$\langle \e^{ \i \vect Q \cdot (\vect R_n - \vect R_{n'})} \rangle \approx 0$. 

To describe the fluctuation of localized moments $\vect J_n$, we adopt a generic Gaussian action given by 
\cite{Moriya1985,Moriya1973a,Moriya1973b,Hertz1976,Millis1993}, 
\begin{equation}
A_{\rm Gauss} = \frac{1}{2}\sum_{\vect p, l}
D^{-1}_{\vect p}( \i \omega_l) \, {\vect J}_{\vect p}( \i \omega_l) \cdot {\vect J}_{-\vect p} (-\i \omega_l)
\label{eq:Agauss}
\end{equation}
with 
\begin{equation}
D_{\vect p} (\i \omega_l) = \frac{1}{\delta + A |\vect p - \vect Q|^2 + |\omega_l|/\Gamma_{\vect p}}. 
\label{eq:Dq}
\end{equation}
Here, $\vect J_{\vect p}( \i \omega_l)$ is a space and imaginary-time $\tau$ Fourier transform of $\vect J_{n}(\tau)$, 
where we made the $\tau$ dependence explicit, and 
$\omega_l = 2 l \pi T$ is the bosonic Matsubara frequency. 
Parameter $A$ is introduced as a constant so that $A |\vect p - \vect Q|^2$ has the unit of energy, and  
$\delta$ is the distance from the magnetic transition temperature and is related to the magnetic correlation length as 
$\xi \propto \delta^{-1/2}$ at $T>T_{\rm N,C}$ and to the ordered magnetic moment as $M(T) \propto \delta^{1/2}$ at $T<T_{\rm N,C}$. 
$\Gamma_{\vect p}$ represents the Landau damping, whose momentum dependence is neglected for AFM systems, $\Gamma_{\vect p} = \Gamma$, since it is weak near the magnetic wave vector $\vect Q$. 
For FM systems with $\vect Q = \vect 0$, the damping term has a momentum dependence as $\Gamma_{\vect p} = \Gamma p$. 
With impurity scattering or disorder, $\Gamma_{\vect p}$ remains finite below a cutoff momentum $|\vect p| \le q_{\rm c}$ as $\Gamma_{\vect p} = \Gamma q_{\rm c}$.

While the above Gaussian action can be derived by solving an interacting electron model, 
it is a highly nontrivial problem and dependent on the detail of theoretical model and the target material. 
Instead, we adopt a conventional approach, where the material dependence is described by a small number of parameters and 
derive the spin Hall conductivity arising from the spin fluctuation and the subleading terms of Eq.~(\ref{eq:Kondo}).  
In fact, theoretical analyses based on this Gaussian action 
have been successful to explain many experimental results on itinerant magnets \cite{Moriya1985}. 
In principle, $\delta$ depends on temperature and is determined by solving self-consistent equations for a full model including non-Gaussian terms 
\cite{Moriya1973a,Moriya1973b,Hertz1976,Moriya1985,Millis1993,Nagaosa1999}. 
However, the temperature dependence of $\delta$ is known for the following three cases in three dimension. 
I: $\delta \propto T-T_{\rm N,C}$ at $T \gtrsim T_{\rm N,C}$, 
II: $\delta \propto M^2(T) \propto T_{\rm N,C}-T$ at $T \lesssim T_{\rm N,C}$, 
and
III: $\delta \propto T^{3/2}$ at $T \sim 0$ when $T_{\rm N} \rightarrow 0$, i.e., approaching the QCP. 
For FM with impurity scattering or disorder $\delta \propto T^{3/2}$ at $T \sim 0$ when $T_{\rm C} \rightarrow 0$, while for clean FM $\delta \propto T^{4/3}$. 

\begin{figure}
\begin{center}
\includegraphics[width=0.9\columnwidth, clip]{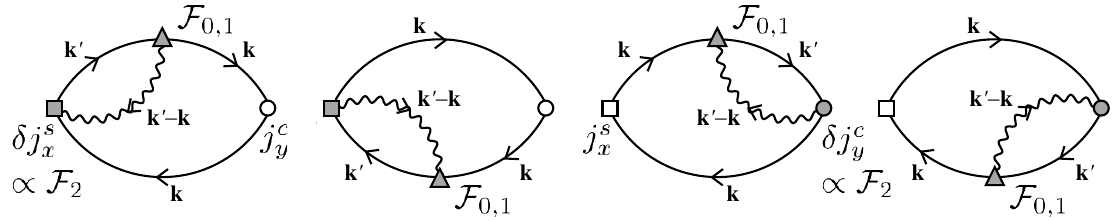}
\caption{Diagrammatic representation for the side-jump contribution. 
Solid (wavy) lines are the electron Green's functions (the SF propagators). 
Squares (circles) are the spin (charge) current vertices, with filled symbols representing the velocity correction with ${\cal F}_2$, 
i.e., side jump. 
Filled triangles are the interaction vertices with ${\cal F}_{0,1}$. 
Adapted from Okamoto, S. Egami, T. \& Nagaosa, N. Phys. Rev. Lett. {\bf 123}, 196603 (2019).}
\label{fig:sidejump}
\end{center}
\end{figure}

\begin{figure}
\begin{center}
\includegraphics[width=0.95\columnwidth, clip]{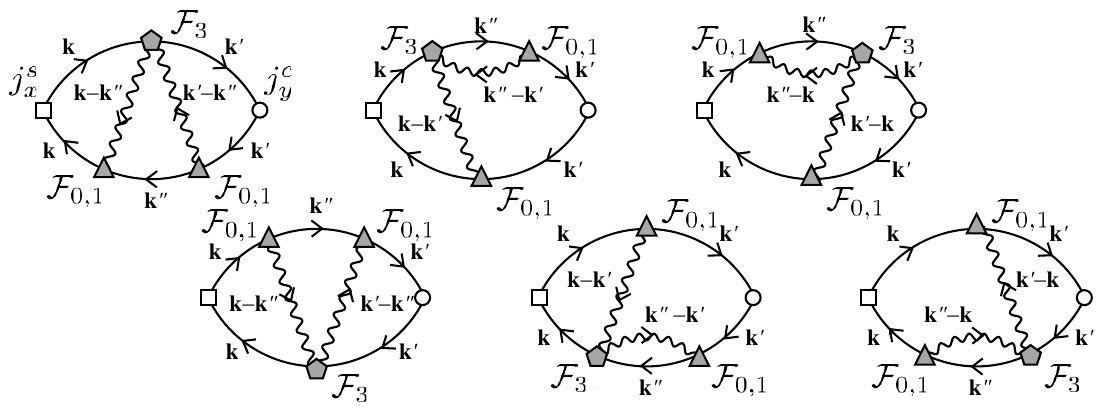}
\caption{Diagrammatic representation for the skew-scattering contribution. 
Filled pentagons are the interaction vertices with ${\cal F}_3$. 
The definitions of the other symbols or lines are the same as in Fig.~\ref{fig:sidejump}. 
Adapted from Okamoto, S. Egami, T. \& Nagaosa, N. Phys. Rev. Lett. {\bf 123}, 196603 (2019).}
\label{fig:skew}
\end{center}
\end{figure}

\subsection*{Spin-Hall conductivity}

With the above preparations, we analyze the SH conductivity using the Matsubara formalism, 
by which one can take the dynamical SF into account via a diagramatic technique.  
%
Here, the frequency-dependent SH conductivity is considered as $\sigma_{\rm SH}(\i \Omega_l)$. 
$\Omega_l$ is the bosonic Matsubara frequency, 
which is analytically continued to real frequency as $\i \Omega_l \rightarrow \Omega + \i 0^+$ at the end of the analysis, 
and then the DC limit, $\Omega \rightarrow 0$, is taken to obtain $\sigma_{\rm SH}$. 
Based on the diagrammatic representations in Figs.~\ref{fig:sidejump} and \ref{fig:skew}, 
$\sigma_{\rm SH}$ is expressed in terms of electron Green's function $G$ and the propagator $D$ of the longitudinal SF. 
While the transverse SFs or spin wave excitations exist below the magnetic transition temperature, 
the scattering of electrons by such SFs does not show a critical behavior  
\cite{Schrieffer1989,Woelfle2021}, and its contribution is expected to be small. 
Therefore, for our analysis, we consider only longitudinal SFs below $T_{\rm N,C}$.

We first focus on the SH effect by the AFM SF. 
By carrying out the Matsubara summations, the energy integrals and the momentum summations as detailed in Supplementary Note 2, we find 
\begin{equation}
\sigma^{\rm side \, jump}_{\rm AFM,SH} \approx \frac{e^2 m \tau_{\vect k}}{2 \pi^3 \hbar^4 |\vect Q|} \, A_{\rm AFM}^{\rm side \, jump} \, \tilde I_{\rm AFM}(T,\delta) 
\label{eq:sigmaAFMSJ}
\end{equation}
for the side-jump contribution and 
\begin{equation}
\sigma^{\rm skew \, scat.}_{\rm AFM,SH} \approx 
 \frac{e^2 \tau_{\vect k}^2 \, \Gamma \, \delta}{2 \pi^2 \hbar^3 |\vect Q|}
\, A_{\rm AFM}^{\rm skew \, scat.} \, I_{\rm AFM}^2 (T,\delta) 
\label{eq:sigmaAFMSS}
\end{equation}
for the skew-scatting contribution. 
Here, $e$ is the elementary charge. 
In both cases, $\tau_{\vect k}$ is the carrier lifetime on the Fermi surface at special momenta $\vect k$ that satisfy the nesting condition. 
Such momenta $\vect k$ form loops on the Fermi surface. 
With the parabolic band, 
the carrier lifetime due to the AFM SF along such loops is independent of momentum as detailed in Supplementary Note 4.
The momentum dependence of the carrier lifetime due to other effects, such as disorder and phonons, is weak. 
Thus, we assume that $\tau_{\vect k}$ is a constant. 
The functions $\tilde I_{\rm AFM}(T,\delta)$ and $I_{\rm AFM}(T,\delta)$ defined in Supplementary Note 2 
represent the coupling between conduction electrons and the dynamical SF. 
$A_{\rm AFM}^{\rm side \, jump}$ and $A_{\rm AFM}^{\rm skew \, scatt.}$ are constants defined by the integrals over the azimuth angle of momentum $\vect k$ measured from the direction of $\vect Q$ as described in Supplementary Note 2. 
Since the angle integrals give only geometrical factors of ${\cal O} (1)$,  
$A_{\rm AFM}^{\rm side \, jump} \approx {\cal F}_0 {\cal F}_2 k_{\rm F}^2$ and
$A_{\rm AFM}^{\rm skew \, scatt.} \approx {\cal F}_0^2 {\cal F}_3 k_{\rm F}^4$. 
%

Similarly, the SH conductivity due to the FM SF is obtained (for details, see Supplementary Note 3) as 
\begin{equation}
\sigma^{\rm side \, jump}_{\rm FM,SH} \approx \frac{2 e^2 \tau_{\vect k}}{m} \, A_{\rm FM}^{\rm side \, jump} \, I_{\rm FM}(T,\delta) 
\label{eq:sigmaFMSJ}
\end{equation}
for the side-jump contribution and 
\begin{equation}
\sigma^{\rm skew \, scat.}_{\rm FM,SH} \approx 
 \frac{4 e^2 \hbar \tau_{\vect k}^2}{m^2}
\, A_{\rm FM}^{\rm skew \, scat.} \, I_{\rm FM}^2(T,\delta) 
\label{eq:sigmaFMSS}
\end{equation}
for the skew-scatting contribution. 
Here, $\tau_{\vect k}$ is the carrier lifetime on the Fermi surface, 
and the constants $A_{FM}^{side \, jump}$ and $A_{FM}^{skew \, scatt.}$ are given by 
$A_{\rm FM}^{\rm side \, jump} \approx \frac{m k_{\rm F}^3}{\pi^2 \hbar} {\cal F}_0 {\cal F}_2$ and 
$A_{\rm FM}^{\rm skew \, scatt.} \approx \frac{m k_{\rm F}^5}{\pi^2 \hbar^2} {\cal F}_0^2 {\cal F}_3$, respectively.  
The function $I_{\rm FM}(T,\delta)$ is defined in Supplementary Note 3. 

\begin{table*}
\caption{$T$ dependence of $\sigma_{\rm SH}$. 
An additional $T$ dependence appears via the carrier lifetime $\tau_{\vect k}$. 
Note that the scaling law breaks down in the vicinity of the transition temperature $T_{\rm N,C}$ 
as indicated by shades in Figs.~\ref{fig:AFMtdepIandII} and \ref{fig:FMtdepIandII}. See the main text for details.}
\begin{tabular}{cc|cccc}
\hline
&&&\\[-0.9em]
\multicolumn{2}{c|}{$T$ regime} & $\sigma_{\rm AFM,SH}^{\rm side \, jump}$ & $\sigma_{\rm AFM,SH}^{\rm skew \, scatt.}$ & $\sigma_{\rm FM,SH}^{\rm side \, jump}$ & $\sigma_{\rm FM,SH}^{\rm skew \, scatt.}$ \\
\hline \hline 
&&&\\[-0.8em]
I & $T>T_{\rm N,C}$ &  
\, $\propto \tau_{\vect k} T^2/\sqrt{T-T_{\rm N}}$ \, & \,  $\propto \tau_{\vect k}^2 T^6/(T-T_{\rm N})$ \,
& \, $\propto \tau_{\vect k} T^3/(T-T_{\rm C})$ \, & \, $\propto \tau_{\vect k}^2 T^6/(T-T_{\rm C})^2$ \, \\
&&&\\[-0.8em]
II & $T<T_{\rm N,C}$ & 
$ \propto \tau_{\vect k} T^2/\sqrt{T_{\rm N}-T}$ & $\propto \tau_{\vect k}^2 T^6/(T_{\rm N}-T)$ 
& $\propto \tau_{\vect k} T^3/(T_{\rm C}-T)$  & $\propto \tau_{\vect k}^2 T^6/(T_{\rm C}-T)^2$ \\
[-0.5em]
\multicolumn{2}{l|}{\hspace{-0.5em} 
\line(1,0){1em} \, \line(1,0){1em} \, \line(1,0){1em} \, \line(1,0){1em} \, \line(1,0){1em} \, \line(1,0){1em} \, \line(1,0){1em}} 
& 
\multicolumn{4}{l}{
\line(1,0){1em} \, \line(1,0){1em} \, \line(1,0){1em} \, \line(1,0){1em} \, \line(1,0){1em} \, \line(1,0){1em} \, 
\line(1,0){1em} \, \line(1,0){1em} \, \line(1,0){1em} \, \line(1,0){1em} \, \line(1,0){1em} \, \line(1,0){1em} \, 
\line(1,0){1em} \, \line(1,0){1em} \, \line(1,0){1em} \, \line(1,0){1em} \, \line(1,0){1em} \, \line(1,0){1em} \, 
\line(1,0){1em} \, \line(1,0){1em} \, \line(1,0){1em} \, \line(1,0){1em} \, \line(1,0){1em} \, \line(1,0){1em} \, 
\line(1,0){1em}
}  \\
III &   $T \sim 0$ ($T_{\rm N,C} \rightarrow 0$) & 
$\propto \tau_{\vect k} T^{3/2}$ & $\propto \tau_{\vect k}^2 T^{9/2}$ 
& $\propto \tau_{\vect k} T^{3/2}$$\dag$  & $\propto \tau_{\vect k}^2 T^3$$\ddag$ \\
\hline \\[-0.8em]
\end{tabular}
\begin{tabular}{rl}
$\dag$&In the absence of disorder or impurity scattering, this temperature dependence is modified as $\propto \tau_{\vect k} T^{5/3}$.$\S$\\
$\ddag$&In the absence of disorder or impurity scattering, this temperature dependence is modified as $\propto \tau_{\vect k}^2 T^{10/3}$$\S$.\\
$\S$&Not considered in the main text, but briefly discussed in Supplementary Note 5. 
\end{tabular}
\label{tab:tableSigmaSH}
\end{table*}

\subsection*{Temperature dependence of the Spin-Hall conductivity}

Reflecting the temperature dependence of spin dynamics, $\sigma_{\rm SH}$ by the SF could show a strong temperature dependence. 
This is governed by the functions $\tilde I_{\rm AFM}(T,\delta)$, $I_{\rm AFM}(T,\delta)$, and $I_{\rm FM}(T,\delta)$, 
and the carrier lifetime $\tau_{\vect k}$. 
$\tau_{\vect k}$ has several contributions, such as  
the disorder or impurity effects $\tau_{\rm dis}$, which $T$ dependence is expected to be small, 
the electron-electron interactions $\tau_{\rm ee}$, the electron-phonon interactions $\tau_{\rm ep}$, 
and the scattering due to the SF $\tau_{\rm sf}$. 
Within the current model, $\tau_{\rm sf} \approx \frac{\hbar}{2} {\cal F}_0^{-2} I_{\rm AFM,FM}^{-1} (T,\delta)$ (see Supplementary Note 4 for details).

In addition to the different momentum dependence in the damping term $\Gamma_{\vect p}$, 
the AFM SF and the FM SF have fundamentally different character due to the momentum conservation during scattering events. 
For the AFM case, electrons scattered by the SF gain or lose momentum $\vect Q$. 
As a result, $\sigma_{\rm AFM,SH}^{\rm skew \, scatt.}$ in Eq.~(\ref{eq:sigmaAFMSS}) has extra $\delta$ 
[for comparison, see Eq.~(\ref{eq:sigmaFMSS})]. 
Furthermore, 
$\sigma_{\rm AFM,SH}^{\rm side \, jump}$ has $\tilde I_{\rm AFM}(\delta, T)$, whose temperature dependence somewhat differs from $I_{\rm AFM}(\delta, T)$. 
$\tilde I_{\rm AFM} (\delta, T)$ has the same temperature dependence of the scattering rate due to the AFM SF 
as reported by Ref. ~\cite{Woelfle2021}. 
While the result of Ref. ~\cite{Woelfle2021} was obtained by loosening the momentum conservation by averaging the electron self-energy over the Fermi surface, 
the momentum dependence is explicitly considered in our $\sigma_{\rm AFM,SH}^{\rm side \, jump}$. 
These differences in the scattering process lead to the different temperature dependence in $\sigma_{\rm SH}$ by the AFM SF and the FM SF.

Supplementary Table~I summarizes 
the $T$ dependence of $\delta$ and $T$-$\delta$ dependence of $\tilde I_{\rm AFM} (T,\delta)$, $I_{\rm AFM} (T,\delta)$, and $I_{\rm FM} (T,\delta)$ 
in three $T$ regimes 
I\,--\,III and in the vicinity of the magnetic phase transition at $T_{\rm N,C}$ between regimes I and II. 
By including the $T$ dependence of $\delta$, the full $T$ dependence of $\tilde I_{\rm AFM} (T,\delta)$ and $I_{\rm AFM,FM} (T,\delta)$ is fixed as follows: 
In the regimes I and II, $\tilde I_{\rm AFM} (T,\delta)$, $I_{\rm AFM}(T,\delta)$, and $I_{\rm FM}(T,\delta)$ are enhanced as $T \rightarrow T_{\rm N,C}$ as 
$\tilde I_{\rm AFM} (T,\delta) \propto \delta^{-1/2} \propto 1/\sqrt{|T-T_{\rm N}|}$, $I_{\rm AFM}(T,\delta) \propto \delta^{-1} \propto 1/|T-T_{\rm N}|$, and 
$I_{\rm FM}(T,\delta) \propto \delta^{-1} \propto 1/|T-T_{\rm C}|$, respectively. 
While the divergence of $\tilde I_{\rm AFM}(T,\delta)$ is cutoff at $T_{\rm N}$, 
the $1/|T-T_{\rm N,C}|$ divergence of $I_{\rm AFM,FM}(T,\delta)$ at $T_{\rm N,C}$ is weakened to the logarithmic divergence $-\ln |T-T_{\rm N}|$ or the smaller power $1/\sqrt{|T-T_{\rm C}|}$. 
Note however that this behavior of $I_{\rm AFM,FM}(T,\delta)$ right at $T_{\rm N,C}$ is a result of the current treatment which does not include the feedback 
between the carrier lifetime and the SF spectrum. 
We anticipate that including such feedback effects will cutoff these divergences. 
Since this requires one to solve the full Hamiltonian, including electron-electron interactions self-consistently, 
such a treatment is left for the future study. 
The SH angle $\Theta_{\rm SH} = \sigma_{\rm SH}/\sigma_{\rm c}$, where $\sigma_c$ is the charge conductivity, is expected to be much smaller than 1, 
even though  $\Theta_{\rm SH}$ could be enhanced at the critical temperature, 
because $\sigma_{\rm SH}$ and $\sigma_{\rm c}$ are both proportional to the carrier lifetime. 
On the other hand, the behavior near the QCP, the regime III, is qualitatively reliable. 
This is because, the scattering rate $\tau_{\vect k}^{-1}$ for the pure case and $I_{\rm AFM,FM}(T,\delta)$ approach $0$ with $T \rightarrow 0$ 
and, and fulfill the self-consistency condition between them. 
In this regime, however, $\tau_{\vect k}$ diverges with $T \rightarrow 0$ without disorder effects. 
This could leads to the pathological divergence of $\sigma_{\rm SH}$. 
We will not consider such a situation in the main text, and give a brief discussion in Supplementary Note 5. 


\begin{figure}
\begin{center}
\includegraphics[width=0.8\columnwidth, clip]{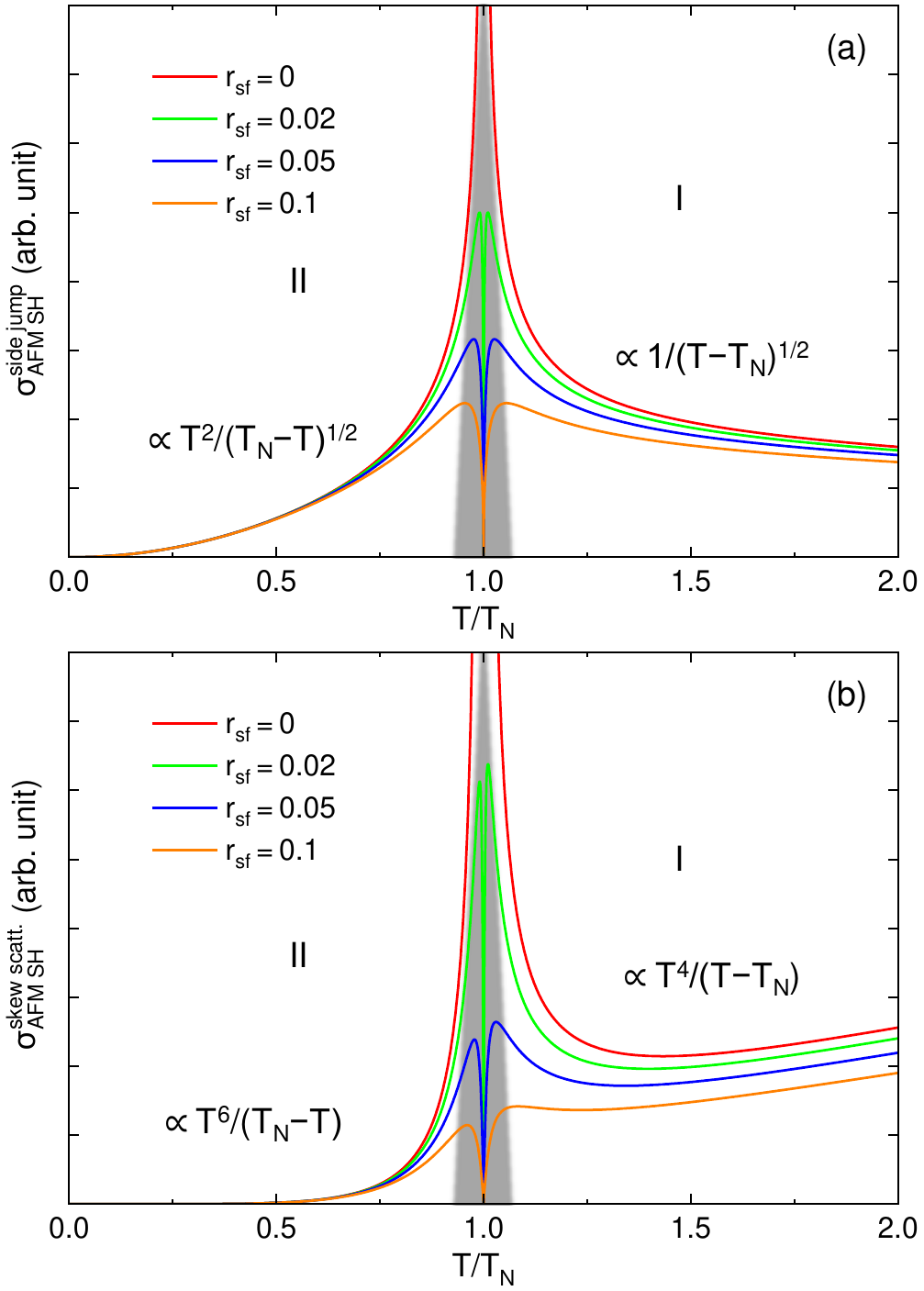}
\caption{Schematic temperature dependence of the SH conductivity of antiferromagnets in the $T$ regimes I and II with nonzero $T_{\rm N}$. 
The side-jump contribution $\sigma_{\rm AFM,SH}^{\rm side \, jump}$ is shown in (a) 
and the skew-scattering contribution $\sigma_{\rm AFM,SH}^{\rm skew \, scatt.}$ is shown in (b). 
The carrier lifetime is modeled as $\tau_{\vect k}^{-1} = r_{\rm dis} + r_{\rm ee} T^2 + r_{\rm sf} T^3/|T-T_{\rm N}|$, with $r_{\rm dis}$, $r_{\rm ee}$, and $r_{\rm sf}$ terms representing the disorder and impurity effects, electron-electron interaction, and the AFM SF, respectively. 
$r_{\rm sf}$ is varied with fixing $r_{\rm dis}=r_{\rm ee}=1$. 
Shaded areas indicate where the current treatment breaks down, requiring the self-consistent treatment. }
\label{fig:AFMtdepIandII}
\end{center}
\end{figure}

\begin{figure}
\begin{center}
\includegraphics[width=0.8\columnwidth, clip]{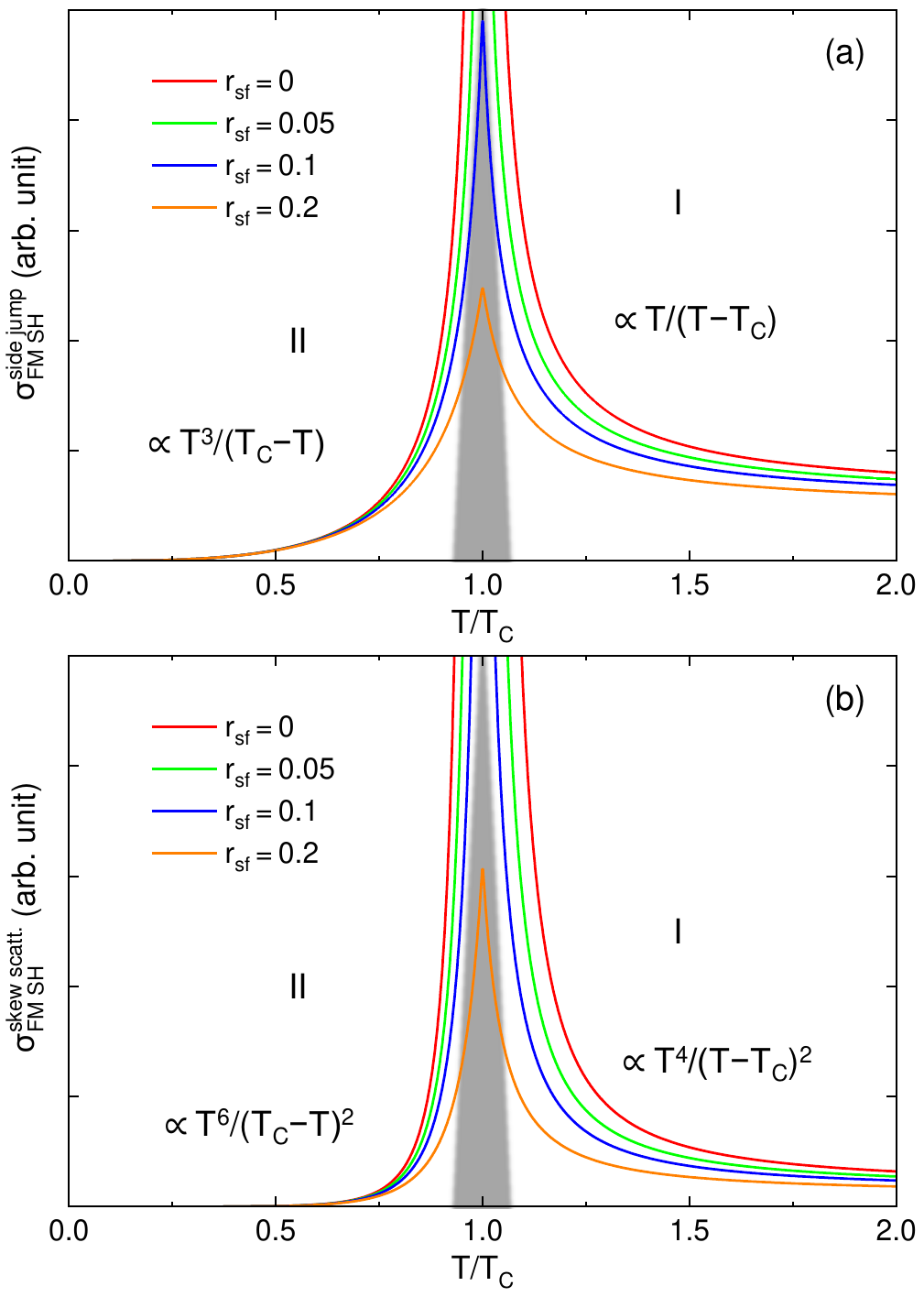}
\caption{Schematic temperature dependence of the SH conductivity ferromagnets in the $T$ regimes I and II with nonzero $T_{\rm C}$. 
The side-jump contribution $\sigma_{\rm FM,SH}^{\rm side \, jump}$ is shown in (a) 
and the skew-scattering contribution $\sigma_{\rm FM,SH}^{\rm skew \, scatt.}$ is shown in (b). 
The carrier lifetime is modeled as $\tau_{\vect k}^{-1} = r_{\rm dis} + r_{\rm ee} T^2 + r_{\rm sf} T^3/|T-T_{\rm C}|$, with $r_{\rm dis}$, $r_{\rm ee}$, and $r_{\rm sf}$ terms representing the disorder and impurity effects, electron-electron interaction, and the FM SF, respectively. 
$r_{\rm sf}$ is varied with fixing $r_{\rm dis}=r_{\rm ee}=1$. 
Shaded areas indicate where the current treatment breaks down, requiring the self-consistent treatment. }
\label{fig:FMtdepIandII}
\end{center}
\end{figure}

The temperature dependence of $\sigma_{\rm SH}$ coming from $\delta$, $\tilde I_{\rm AFM}$, and $I_{\rm AFM,FM}$
is summarized in Table~\ref{tab:tableSigmaSH}. 
Reflecting the diverging behavior of $\tilde I_{\rm AFM}$ and $I_{\rm AFM,FM}$, 
$\sigma_{\rm SH}$ is sharply enhanced as $T \rightarrow T_{\rm N,C}$ in the regimes I and II,  
as displayed in Figures~\ref{fig:AFMtdepIandII} for the AFM case and \ref{fig:FMtdepIandII} for the FM case. 
Here, the approximate inverse carrier lifetime appropriate in these $T$ regimes is considered as 
$\tau_{\vect k}^{-1} = r_{\rm dis} + r_{\rm ee} T^2 + r_{\rm sf} T^3/|T-T_{\rm N,C}|$, 
where $r_{\rm dis}$, $r_{\rm ee}$, and $r_{\rm sf}$ terms correspond to the disorder effect, electron-electron interaction \cite{Baber1937}, 
and the SF, respectively. 
Focusing on the low $T$ behavior, we ignored the electron-phonon coupling, which would contribute to the carrier lifetime 
at high temperatures close to the Debye temperature \cite{Bloch1930,Ziman1960}. 
%
In the current theory, there are three energy units; the Fermi energy $\varepsilon_{\rm F} \approx \hbar v_{\rm F}/a$, the spin stiffness $A/a^2$
and magnetic transition temperature $T_{\rm N,C}$. 
The temperature dependence of $\sigma_{\rm AFM,FM}^{\rm skew \, scatt.}$ and $\sigma_{\rm FM}^{\rm side \, jump}$ is from $I_{\rm AFM,FM}(T,\delta)$, and therefore $T$ is scaled by $\varepsilon_{\rm F}$, while 
$\sigma_{\rm AFM}^{\rm side jump}$ is from $\tilde I_{\rm AFM}(T,\delta)$ and $T$ is scaled by $A/a^2$. 
For the analytical plots, we use the dimensionless unit for temperature, where $T$ is scaled by these energy units, 
and $T_{\rm N,C}=1$ for the $T$ regimes I and II. 
With this convention, $r_{\rm dis}$, $r_{\rm ee}$, $r_{\rm sf}$ have the unit of inverse time.  

Despite the diverging trend as $T \rightarrow T_{\rm N}$, 
$\sigma_{\rm AFM,SH}^{\rm side \, jump}$ and $\sigma_{\rm AFM,SH}^{\rm skew \, scatt.}$ sharply drop to zero in the vicinity of $T_{\rm N}$ 
with nonzero $ r_{\rm sf}$. 
This is caused by the suppression of $\tau_{\vect k}$ due to the SF
as $\tau_{\vect k} \approx r_{\rm sf}^{-1} |T-T_{\rm N}|/T^3$. 
We anticipate that a self-consistent treatment of the original interacting electron  model kills this entire suppression, leading to a smooth $T$ dependence of $\sigma_{\rm SH}$. 

$\sigma_{\rm FM,SH}^{\rm side \, jump}$ and $\sigma_{\rm FM,SH}^{\rm skew \, scatt.}$ have stronger $T$ dependence than the AFM counterparts, 
leading to the divergence with $T$ approaches $T_{\rm C}$ with $r_{\rm sf}=0$. 
Nonzero $ r_{\rm sf}$ suppresses the divergence in $\sigma_{\rm FM,SH}^{\rm side \, jump}$ and $\sigma_{\rm FM,SH}^{\rm skew \, scatt.}$ 
in the vicinity of $T_{\rm C}$, leading to sharp cusps. 
However, similar to the AFM case, we anticipate that a self-consistent treatment of the original model leads to a smooth $T$ dependence of $\sigma_{\rm SH}$ across $T_{\rm C}$. 

Because of the competition between the divergence of $I_{\rm AFM,FM}(T,\delta)$ and the suppression of $\tau_{\vect k}$, 
it might be challenging to deduce the precise temperature scaling of the SH conductivity at $T_{\rm N,C}$. 
Nevertheless, our result summarized in Table~\ref{tab:tableSigmaSH} will be helpful to analyze experimental SH conductivity because the two contributions are separated. 

For the quantum critical regime III, the carrier lifetime has the temperature dependence as 
$\tau_{\vect k}^{-1} = r_{\rm dis} + r_{\rm ee} T^2 + r_{\rm sf} T^{3/2}$ for the AFM case and the FM with disorder effects. 
The temperature dependence of $\sigma_{\rm SH}$ is strongly influenced by that of $\tau_{\vect k}$. 
Thus, here we discuss the cases with disorder effect which make $\tau_{\vect k}$ finite at $T=0$. 
Special cases, where the disorder effect is absent and $\tau_{\vect k}$ becomes infinity at $T=0$, will be briefly discussed in Supplementary Note 5. 

\begin{figure}
\begin{center}
\includegraphics[width=0.8\columnwidth, clip]{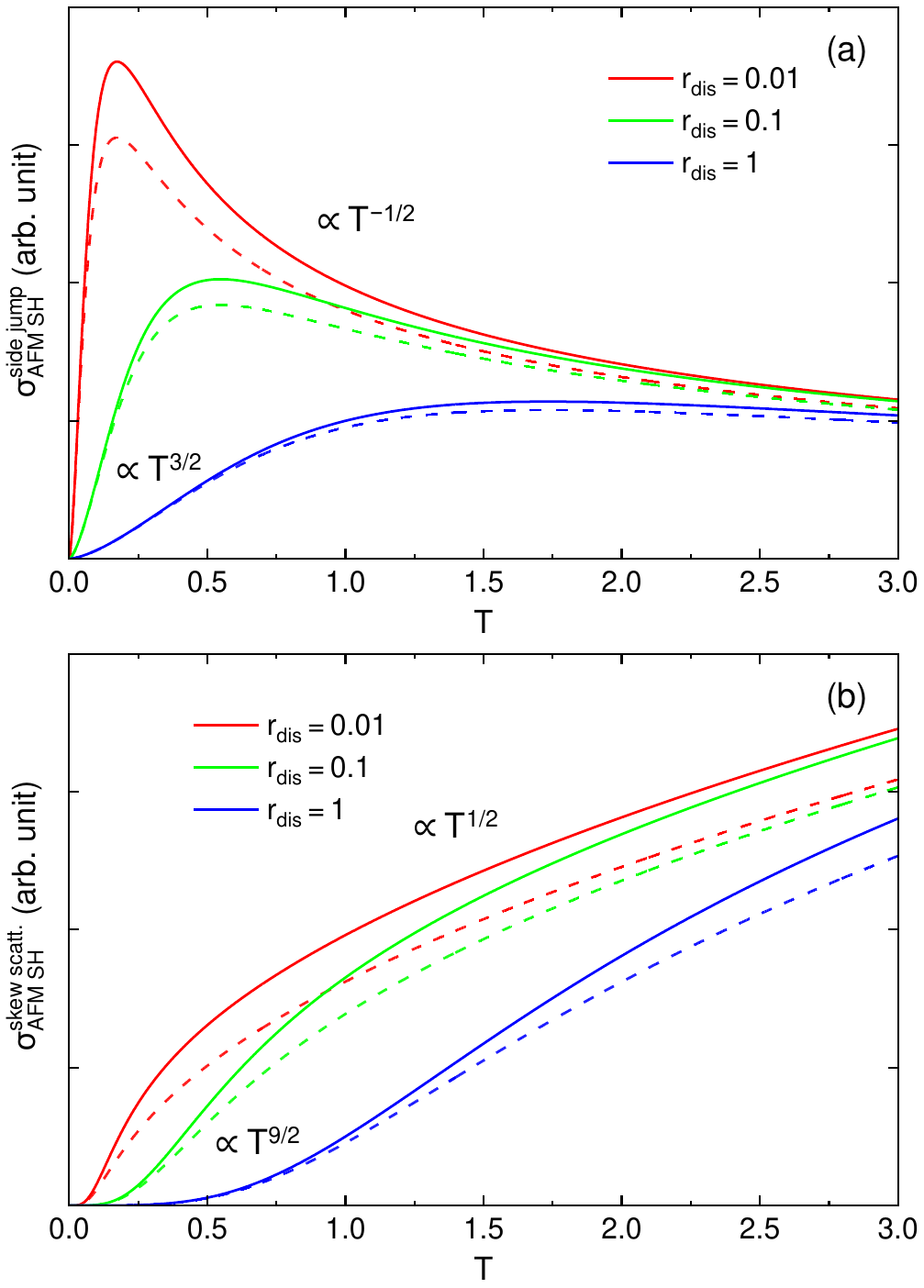}
\caption{Schematic temperature dependence of the SH conductivity of antiferromagnets in the $T$ regime III with $T_{\rm N} \rightarrow 0$. 
The side-jump contribution $\sigma_{\rm AFM,SH}^{\rm side \, jump}$ is shown in (a), 
and the skew-scattering contribution $\sigma_{\rm AFM,SH}^{\rm skew \, scatt.}$ shown in (b). 
The carrier lifetime is modeled as $\tau_{\vect k}^{-1} = r_{\rm dis} + r_{\rm ee} T^2 + r_{\rm sf} T^{3/2}$, and 
$r_{\rm dis}$ is varied with fixing $r_{\rm ee}=1$ and $r_{\rm sf}=0.0$ $(0.1)$ for solid (dashed) lines.}
\label{fig:AFMtdepIII}
\end{center}
\end{figure}

\begin{figure}
\begin{center}
\includegraphics[width=0.8\columnwidth, clip]{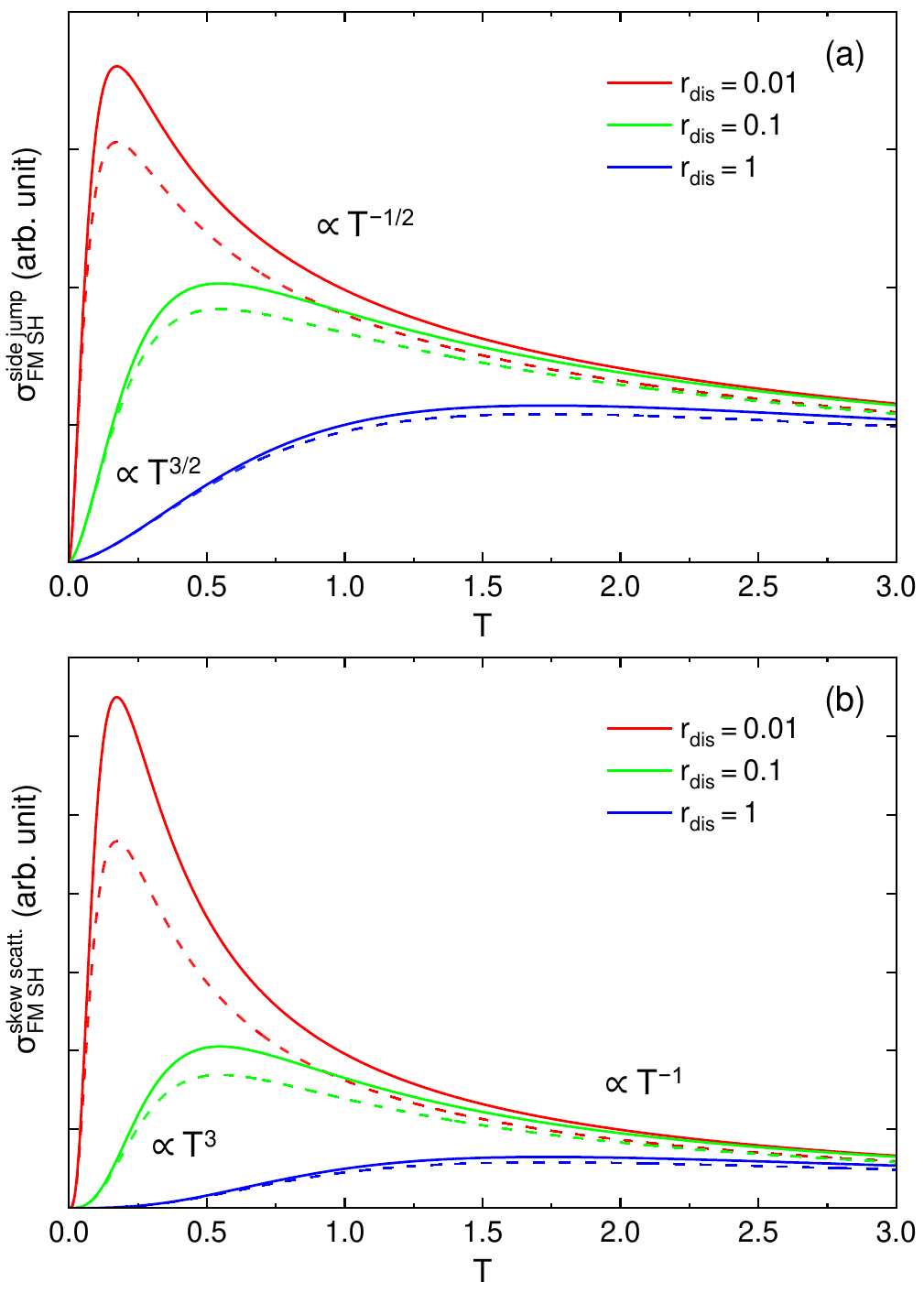}
\caption{Schematic temperature dependence of the SH conductivity of ferromagnets in the $T$ regime III with $T_{\rm C} \rightarrow 0$. 
The side-jump contribution $\sigma_{\rm FM,SH}^{\rm side \, jump}$ is shown in (a), 
and the skew-scattering contribution $\sigma_{\rm FM,SH}^{\rm skew \, scatt.}$ shown in (b). 
The carrier lifetime is modeled as $\tau_{\vect k}^{-1} = r_{\rm dis} + r_{\rm ee} T^2 + r_{\rm sf} T^{3/2}$, and 
$r_{\rm dis}$ is varied with fixing $r_{\rm ee}=1$ and $r_{\rm sf}=0$ $(0.1)$ for solid (dashed) lines.}
\label{fig:FMtdepIII}
\end{center}
\end{figure}

The schematic temperature dependence of $\sigma_{\rm AFM,SH}$ and $\sigma_{\rm FM,SH}$ is shown in Fig.~\ref{fig:AFMtdepIII} and \ref{fig:FMtdepIII}, respectively. 
With nonzero $r_{\rm dis}$, all $\sigma_{\rm SH}$ approach 0 with $T$ goes to 0 but with different $T$ scaling; 
$\sigma_{\rm AFM,SH}^{\rm side \, jump} \propto T^{3/2}$, $\sigma_{\rm AFM,SH}^{\rm skew \, scatt.} \propto T^{9/2}$,  
$\sigma_{\rm FM,SH}^{\rm side \, jump} \propto T^{3/2}$, and $\sigma_{\rm FM,SH}^{\rm skew \, scatt.} \propto T^{3}$. 
For the latter two cases, different power laws of $T$ were predicted in Ref.~\cite{Okamoto2019} as described in Supplementary Note 3. 

Interestingly, $\sigma_{\rm AFM,SH}^{\rm side \, jump}$ and $\sigma_{\rm FM,SH}^{\rm side \, jump}$ show formally the same leading $T$ dependence 
because the divergence of  the SF propagator has a cutoff by $|\vect Q|$ in the former and $q_{\rm c}$ in the latter. 
On the other hand, 
$\sigma_{\rm AFM,SH}^{\rm skew \, scatt.}$ and $\sigma_{\rm FM,SH}^{\rm skew \, scatt.}$ show contrasting $T$ dependence; 
the former continuously decreases with decreasing $T$ while the latter first increases, shows maximum, and finally goes to zero with decreasing $T$ 
because of the competition between $\tau_{\vect k}$ and $I_{\rm FM}(T,\delta)$. 

\section*{D\lowercase{iscussion}} 

We have seen that the SH effect in magnetic metallic systems induced by spin fluctuations has different contributions with different temperature scaling. 
In this section, we consider remaining questions regarding the relative strength between different contributions as well as the experimental/materials realization of our theory. 

\vspace{0.5em}
%
When the FM critical fluctuation is dominant in the carrier lifetime in the temperature regime III, 
the carrier lifetime is given by 
$\tau_{\rm sf} \approx 
 \hbar / 2  \bigl\{ F_0 + 2 F_1 k_{\rm F}^2 \bigr\}^2  I_{\rm FM} (T,\delta)$, and therefore 
the ratio between the maximum $\sigma_{\rm FM,SH}^{\rm skew \, scatt.}$ and the maximum $\sigma_{\rm FM,SH}^{\rm side \, jump}$ is estimated 
as 
$\sigma_{\rm FM,SH}^{\rm skew \, scatt.}/\sigma_{\rm FM,SH}^{\rm side \, jump} \approx \varepsilon_{\rm F} {\cal F}_3/{\cal F}_0 {\cal F}_2$ \{see, 
Eqs.~(\ref{eq:sigmaFMSJ}) and (\ref{eq:sigmaFMSS}) and the discussion in Ref.~\cite{Okamoto2019}\}. 
From the typical interaction strengths ${\cal F}$s and the Fermi energy $\varepsilon_{\rm F}$, the maximum of 
$\sigma_{\rm FM,SH}^{\rm skew \, scatt.}$ is expected to be 1 to 2 orders of magnitude larger than that of $\sigma_{\rm FM,SH}^{\rm side \, jump}$.  
This relation is expected be hold for the FM case with finite $T_{\rm C}$. 

When disorder effects or electron-electron scattering becomes dominant in the carrier lifetime, 
the magnitude of $\tau_{\vect k}$ and $I_{\rm FM}(T,\delta)$ has to be explicitly considered. 
Using the asymptotic form of $I_{\rm FM}(T,\delta)$ near $\delta \sim 0$ in the $T$ regimes of I, II, and III, $I_{\rm FM} (T,\delta) \approx \frac{1}{8 \pi} (\frac{a T}{\hbar v_{\rm F}})^3 \frac{1}{\delta}$ (see Supplementary Note 3 for details)
the ratio between $\sigma^{\rm skew \, scatt.}_{\rm FM,SH}$ and $\sigma^{\rm side \, jump}_{\rm FM,SH}$ is estimated as 
\begin{eqnarray}
\frac{\sigma^{\rm skew \, scatt.}_{\rm FM,SH}}{\sigma^{\rm side \, jump}_{\rm FM,SH}}
\approx 
\frac{k_{\rm F}^2}{4 \pi m} \tau_{\vect k}  {\cal F}_0 \frac{{\cal F}_3}{{\cal F}_2} 
\biggl(\frac{aT}{\hbar v_{\rm F}} \biggr)^3 \frac{1}{\delta}. 
\end{eqnarray}
Because of the factor of $1/\delta$, this ratio diverges when $\delta$ goes to zero as $T$ approaches $T_{\rm C}$ as long as $\tau_{\vect k}$ is finite. 
Thus, the skew-scattering mechanism is expected to become dominant near the critical temperature. 
On the other hand near the FM QCP, the side-jump contribution may grow with lowering temperature when 
the carrier lifetime is dominated by other mechanisms than the SF. 

In the AFM-fluctuation case, the situation is more complicated. 
This is because the side-jump contribution and the skew-scattering contribution have different temperature dependence, 
$\tilde I_{\rm AFM}(T,\delta)$ vs. $I_{\rm AFM}(T,\delta)$, 
while they show similar enhancement near the magnetic transition temperature. 
Therefore, the microscopic parameters determining the SF come into play. 
To see this, first consider the temperature regimes I and II, 
where the leading temperature dependence of $\sigma_{\rm AFM,SH}^{\rm side \, jump}$ and $\sigma_{\rm AFM,SH}^{\rm skew \, scatt.}$ is given by 
\begin{eqnarray}
\sigma_{\rm AFM,SH}^{\rm side \, jump} \approx \frac{e^2 m}{32 \pi^2 \hbar^4} \frac{\tau_{\vect k}}{|\vect Q|} \, 
{\cal F}_0 k_{\rm F}^2 {\cal F}_2\, \frac{T^2}{\Gamma \sqrt{\delta (A/a^2)^3}}
\end{eqnarray}
and 
\begin{eqnarray}
\sigma_{\rm AFM,SH}^{\rm skew \, scatt.} \approx \frac{e^2}{128 \pi^3 \hbar^3} \frac{\tau_{\vect k}^2}{|\vect Q|}\,
{\cal F}_0^2 k_{\rm F}^2 {\cal F}_3\, \biggl( \frac{T}{\varepsilon_{\rm F}} \biggr)^{\!\!6}\frac{\Gamma}{\delta}, 
\end{eqnarray}
respectively. 
Here, we approximate $k_{\rm F} \approx 1/a$ (inverse lattice constant), so that 
$\hbar v_{\rm F}/a \approx \varepsilon_{\rm F}$. 
The ratio between these two contributions leads to 
\begin{eqnarray}
\frac{\sigma_{\rm AFM,SH}^{\rm skew \, scatt.}}{\sigma_{\rm AFM,SH}^{\rm side \, jump}} \approx
\frac{1}{2\pi} \frac{\tau_{\vect k}}{\hbar} \, \Gamma^2 {\cal F}_0 \frac{{\cal F}_3}{{\cal F}_2} \frac{T^4}{\varepsilon_{\rm F}^5} 
\sqrt{\frac{(A/a^2)^3}{\delta}}. 
\end{eqnarray}
Thus, the relative strength depends on both electronic properties and the SF. 
On the electronic part, ({\it i}) longer lifetime $\tau_{\vect k}$, ({\it ii}) larger ${\cal F}_3$ than ${\cal F}_2$, 
and ({\it iii}) smaller Fermi energy $\varepsilon_{\rm F}$ prefer the skew-scattering mechanism over the side jump. 
On the SF part, 
({\it iv}) larger $A$, corresponding to spin stiffness or magnetic exchange, 
({\it v}) larger damping ratio $\Gamma$, which is a dimensionless parameter here but 
is proportional to the electron density of states at the Fermi level,  
and ({\it vi}) smaller $\delta$ prefer the skew-scattering contribution.

Near the AFM QCP (the $T$ regime III), $\sigma_{\rm AFM, SH}^{\rm side \, jump}$ is modified as 
\begin{eqnarray}
\sigma_{\rm AFM,SH}^{\rm side \, jump} \approx \frac{e^2 m}{2 \pi^3 \hbar^4} \frac{\tau_{\vect k}}{|\vect Q|} \, 
{\cal F}_0 k_{\rm F}^2 {\cal F}_2\, \frac{T^{3/2}}{\Gamma \sqrt{(A/a^2)^3}}.
\end{eqnarray}
Hence, with $\delta=T^{3/2}$, the ratio between the two contributions becomes  
\begin{eqnarray}
\frac{\sigma_{\rm AFM,SH}^{\rm skew \, scatt.}}{\sigma_{\rm AFM,SH}^{\rm side \, jump}} \approx
\frac{\pi}{32} \frac{\tau_{\vect k}}{\hbar} \, \Gamma^2 {\cal F}_0 \frac{{\cal F}_3}{{\cal F}_2} \frac{T^3}{\varepsilon_{\rm F}^5} 
\sqrt{(A/a^2)^3}. 
\end{eqnarray}
Thus, $\sigma_{\rm AFM,SH}^{\rm side \, jump}$ is expected to become progressively dominant as $T \rightarrow 0$. 
This could be seen in the contrasting $T$ dependence of $\sigma_{\rm AFM,SH}$ as plotted in Fig.~\ref{fig:AFMtdepIII}. 

\vspace{0.5em}
In this work, we first considered the SH effect by the AFM SF, 
which is relevant to AFM metallic Cr. 
As early studies have reported, metallic Cr shows large SH effect \cite{Du2014,Qu2015}. 
With the small SOC for a $3d$ element, this indicates additional contributions to the SH effect. 
Recent study used high-quality single crystal of Cr and revealed the detailed temperature dependence of the SH conductivity
\cite{Fang2023}. 
Their electric resistivity data does not show a strong anomaly at $T_{\rm N}$. 
This indicates that the carrier lifetime is influenced by magnetic ordering and the AFM SF only weakly 
and, thus, the system is in the perturbative regime, corresponding to very small $r_{\rm sf}$ in the plots of Fig.~\ref{fig:AFMtdepIandII}. 
Thus, the strong enhancement in the SH conductivity could be ascribed to the mechanisms developed in this work. 
The remaining question is 
which mechanism provides the main contribution to the SH effect in Cr, the side-jump mechanism or the skew-scattering mechanism. 
This will be answered when the SF fluctuation spectrum is carefully analyzed. 
Such analyses will also be helpful to understand and predict other AFM metallic systems for the SH effect.

At this moment, we are unaware of experimental reports of the SH effect in the vicinity of the AFM QCP. 
It might be worth investing the SH effect using CeCu$_{6-x}$Au$_x$ \cite{Lohneysen1996} and other Ce compounds \cite{Lohneysen2007}. 
The temperature dependence of the SH conductivity might provide further insight into the nature of their QCP.

The SH effect near the FM critical temperature appears to depend on the material. 
Early studies on Ni-Pd alloys \cite{Wei2012} reported that the temperature dependence of the SH effect is analogous to that of 
the uniform second-order nonlinear susceptibility $\chi_2$, with a positive peak above $T_{\rm C}$ and a negative peak below $T_{\rm C}$, 
thus showing the sign change across $T_{\rm C}$. 
Such a behavior is qualitatively reproduced by a theoretical work by Gu et al. \cite{Gu2012}, 
which adopted a static mean field approximation to the Kondo's model \cite{Kondo1962}. 
The current work, on the other hand, predicts that the SH effect of FM metals is maximized at $T_{\rm C}$, while the same model is used as Gu et al. 
An experimental study by Ou et al. reported the sharp enhancement of the SH effect near $T_{\rm C}$ of Fe-Pt alloys \cite{Ou2018}, 
the behavior resembles our prediction. 
A more recent experimental study by Wu et al. also reported a similar but weaker enhancement of the SH effect of Ni-Cu alloys \cite{Wu2022}. 
How the SH effect depends on the material, changing sign or maximizing at $T_{\rm C}$, remains an open question. 
One possible scenario is that the spin dynamics in Ni-Pd alloys is `classical' in nature, while that in Fe-Pt and  Ni-Cu alloys is more `quantum', 
so that the theoretical analysis presented in this work is more relevant to the latter. 
Detailed experimental analysis on the spin dynamics of these FM metallic alloys using inelastic neutron scattering 
would settle this issue.

It is not obvious which SF generates the larger spin Hall effect, AFM or FM, because the detail of the materials property is involved. 
From the leading temperature dependence, the FM SF gives a stronger temperature dependence of the spin Hall conductivity 
when approaching $T_{\rm C}$ from higher temperature than the AFM fluctuation when approaching $T_{\rm N}$. 
For FM metals, the spin Hall effect is expected to be replaced by the anomalous Hall effect below $T_{\rm C}$, 
which is not the scope of the current study, 
while for AFM metals the spin Hall effect should persist down to low temperatures. 
Thus, both systems could be useful for the spintronic application depending on the temperature range. 

\vspace{0.5em}

To summarize, we developed the comprehensive theoretical description of the spin Hall effect in magnetic metallic systems due to the spin fluctuation. 
The special focus is paid to the antiferromagnetic spin fluctuation with nonzero N{\'e}el temperature $T_{\rm N}$ and $T_{\rm N}=0$, and the FM SF with nonzero Curie temperature $T_{\rm C}$. 
%
In contrast to the spin Hall effect due to the ferromagnetic critical fluctuation, 
where the skew-scattering mechanism is one or two orders of magnitude stronger than the side-jump mechanism, 
the relative strength of the mechanisms 
could be altered depending on the detail of the spin-fluctuation spectrum and temperature. 
In particular, for antiferromagnetic metals, the skew-scattering mechanism becomes progressively dominant 
when approaching the magnetic transition temperature $T_{\rm N}$, 
while the side-jump contribution becomes dominant by lowering temperature below $T_{\rm N}$. 
The crossover from the skew scattering to the side jump also appears in a quantum critical system, where $T_{\rm N}$ is tuned to zero temperature.
Aside from the absolute magnitude of the spin Hall conductivity, antiferromagnetic metals and ferromagnetic metals could be 
complementary in nature. 
This work thus provides an important component in antiferromagnetic spintronics \cite{Balz2018}. 
Many magnetic metallic systems have been reported to show a variety of Hall effects, 
for example the anomalous Hall effect in Fe$_3$GeTe$_2$ \cite{Kato2022}, 
induced by the nontrivial band topology due to the orbital complexity, the spin-orbit coupling, as well as magnetic ordering. 
In the presence of such complexities, the predicted scaling law of the spin Hall effect could be modified, 
while the critical enhancement would not be entirely eliminated. 
Such an interplay will be an exciting research area, but left for the future study.


\section*{D\lowercase{ata availability}}
The data that support the findings of this study are available from the corresponding author upon reasonable request.

\section*{A\lowercase{cknowledgements}}

The research by S.O. was supported by the  U.S. Department of Energy, Office of Science, Basic Energy Sciences, Materials Sciences and Engineering Division. 
N.N. was supported by JST CREST Grant Number JPMJCR1874, Japan, and JSPS KAKENHI Grant number 18H03676. 
We thank Dr. Chi Fang, Prof. Yuan Lu, and Prof. Xiufeng Han for their stimulating discussions and sharing their experimental data. 

Copyright  notice: This  manuscript  has  been  authored  by  UT-Battelle, LLC under Contract No. DE-AC05-00OR22725 with the U.S.  Department  of  Energy.   
The  United  States  Government  retains  and  the  publisher,  by  accepting  the  article  for  publication, 
acknowledges  that  the  United  States  Government  retains  a  non-exclusive, paid-up, irrevocable, world-wide license to publish 
or reproduce the published form of this manuscript, or allow others to do so, for United States Government purposes.  
The Department of Energy will provide public access to these results of federally sponsored  research  in  accordance  with  the  DOE  Public  Access  Plan 
(http://energy.gov/downloads/doe-public-access-plan). 

\section*{A\lowercase{uthor Contributions}}

S.O. conceived the project and carried out calculations with input from N.N. 
Both authors contributed to the writing of the manuscript. 

\section*{C\lowercase{ompeting Interests}}

The authors declare no competing interests.

\vspace{1em}
{\bf Correspondence} and requests for materials should be addressed to S.O. (okapon@ornl.gov).

\clearpage

\onecolumngrid

\begin{center}
{\large \bf Supplementary material: Critical enhancement of the spin Hall effect by spin fluctuations}\\
\vspace{1em}
Satoshi Okamoto,$^1$ and Naoto Nagaosa$^{2}$\\
\vspace{0.5em}
{\small \it $^1$Materials Science and Technology Division, Oak Ridge National Laboratory, Oak Ridge, Tennessee 37831, USA}\\
{\small \it $^2$RIKEN Center for Emergent Matter Science (CEMS), Wako, Saitama 351-0198, Japan}\\
\end{center}

\renewcommand{\figurename}{Supplementary Figure}
\renewcommand{\tablename}{Supplementary Table}
\renewcommand{\thesubsection}{Supplementary Note \arabic{subsection}}
\renewcommand{\thesubsubsection}{Supplementary Note \arabic{subsection}.\arabic{subsubsection}}
\renewcommand{\bibsection}{\subsection*{Supplementary References}}

\setcounter{secnumdepth}{3}

\setcounter{equation}{0}
\setcounter{figure}{0}

\subsection{Kondo model in the momentum space}
\label{subsection:Kondo}

The original model, that was proposed by Kondo in Ref.~\cite{SKondo1962} and adopted by the current authors in Ref.~\cite{SOkamoto2019}, 
has a mixed representation of momentum of conduction electrons and real-space coordinate of localized magnetic moments. 
This is suited to describe metallic systems, where localized magnetic moments do not necessarily form a periodic lattice. 
For magnetic metallic systems, wher localized magnetic moments form a periodic lattice and conduction electrons hop through the same lattice sites, 
it is more convenient to start from a model Hamiltonian represented in the momentum space. 

The model Hamiltonian adopted in Ref.~\cite{SOkamoto2019} is given by 
\begin{eqnarray}
H_{\rm K} \!\!&=&\!\!
-\frac{1}{N} \sum_{n}^{N_{\rm m}} \sum_{\vect k, \vect k'} \sum_{\nu, \nu'} \e^{\i (\vect k' -\vect k) \cdot \vect R_n} 
a_{\vect k \nu}^\dag a_{\vect k' \nu'} 
\biggl[ 2(\vect J_n \cdot \vect s_{\nu \nu'})
\bigl\{ {\cal F}_0 + 2 {\cal F}_1(\vect k \cdot \vect k') \bigr\} 
+ \i {\cal F}_2 \vect J_n \cdot (\vect k' \times \vect k) \nonumber \\
&+&\!\! \i {\cal F}_3 \Bigl\{( \vect J_n \cdot \vect s_{\nu \nu'}) \bigl( \vect J_n \cdot (\vect k' \times \vect k) \bigr) 
+\bigl( \vect J_n \cdot (\vect k' \times \vect k) \bigr) (\vect J_n \cdot \vect s_{\nu \nu'}) 
- \frac{2}{3} (\vect J_n \cdot \vect J_n) \bigl( \vect s_{\nu \nu'} \cdot (\vect k' \times \vect k) \bigr) \Bigr\} \biggr] . 
\label{eq:Kondo}
\end{eqnarray}
Here, $a_{\vect k \nu}^{(\dag)}$ is the annihilation (creation) operator of a conduction electron with momentum $\vect k$ and spin $\nu$, 
$\varepsilon_{\vect k} = \frac{\hbar^2 k^2}{2m}- \varepsilon_{\rm F}$ is the dispersion relation measured from the Fermi energy $\varepsilon_{\rm F}$ 
with the carrier effective mass $m$, 
$\vect s_{\nu \nu'}=\frac{1}{2} \vecs \sigma_{\nu \nu'}$is  the conduction electron spin with the Pauli matrices $\vecs \sigma$,
and  $N (N_{\rm m})$ is the total number of lattice sites (local moments). 
$\vect J_n$ is a local magnetic moment at position $\vect R_n$, when the SOC is weaker than the crystal field splitting and could be treated as a perturbation, 
or the local total angular momentum, when the SOC is strong so that  the total angular momentum is a constant of motion.  
Parameters ${\cal F}_l$ are related to $F_l$ defined in Ref.~\cite{SKondo1962} 
as discussed in Ref.~\cite{SOkamoto2019}. 

For the current periodic system with $N_{\rm m}=N$, we introduce the Fourier transform to localized moments as 
\begin{eqnarray}
\vect J_n = \frac{1}{\sqrt{N}} \sum_{\vect p}  \vect J_{\vect p} \e^{\i \vect p \cdot \vect R_n}
\end{eqnarray}
and the normalization $\frac{1}{N} \sum_n \e^{\i (\vect k -\vect k')\cdot \vect R_n} = \delta_{\vect k \vect k'}$. 
The annihilation operator of an electron $a_{\vect k \nu}$ follows the same rule. 

After introducing this Fourier transform, the Hamiltonian describing the coupling between localized moments and itinerant electrons is 
given entirely in the momentum space as presented in Eq.~(1) in the main text. 

As in the conventional SH effect, a side-jump-type contribution to the SH effect arises from the anomalous velocity. 
Using a standard derivation $\vect v = (\i /\hbar) [H_{\rm K},\vect r]$, the velocity operator is obtained as 
\begin{equation}
\vect v 
=\sum_{\vect k}\frac{\hbar \vect k}{m} a_{\vect k \nu}^\dag a_{\vect k \nu} 
- \frac{\i {\cal F}_2}{\hbar \sqrt{N}}  \sum_{\vect k, \vect k'} \sum_{\nu, \nu'} 
 \vect J_{\vect k - \vect k'}  
\times (\vect k'-\vect k) a_{\vect k \nu}^\dag a_{\vect k' \nu'}. 
\label{eq:velocity}
\end{equation}
Here, we neglected terms involving ${\cal F}_{1,3}$ because these terms 
do not contribute to $\sigma_{\rm SH}$ at the lowest order. 
The second term involving ${\cal F}_{2}$ is the anomalous velocity. 
The charge current and the spin current are given by using the velocity operator as 
$\vect j^{\rm c} = -e \vect v$ and $\vect j^{\rm s} = - e \{\frac{1}{N} \sum_{\vect k} s_{\nu \nu'}^z a_{\vect k \nu}^\dag a_{\vect k \nu'}, \vect v\}$, respectively, 
with $e$ being the elementary charge.  
These current operators have the same dimension. 

\subsection{Spin-Hall conductivity of antiferromagnets}
\label{subsection:SHconductivity}

Here, we highlight the formalism to compute the spin-Hall conductivity $\sigma_{\rm SH}$ as described in Ref.~\cite{SOkamoto2019}. 
The frequency-dependent SH conductivity is formally expressed as 
\begin{eqnarray}
\sigma_{\rm SH}(\i \Omega_l) = \frac{\i}{\i \Omega_l V} \int_0^{1/T} \hspace{-1em} d\tau \e^{\i \Omega_l \tau} 
\langle T_\tau j_x^{\rm s} (\tau) j_y^{\rm c} (0) \rangle. 
\label{eq:sigmaSH}
\end{eqnarray}
Here, $\Omega_l$ is the bosonic Matsubara frequency, 
which is analytically continued to real frequency as $\i \Omega_l \rightarrow \Omega + \i 0^+$ at the end of the analysis, 
and then the DC limit, $\Omega \rightarrow 0$, is taken to obtain $\sigma_{\rm SH}$. 
$V$ is the volume of the system. 
The imaginary time $\tau$ dependence is explicitly shown for the current operators $j_x^{\rm s} (\tau)$ and  $j_y^{\rm c} (0)$. 
These current operators are defined in the main text. 

Based on the diagrammatic representations in Figs.~2 and 3 in the main text, 
$\sigma_{\rm SH}$ is expressed in terms of electron Green's function $G$ and the propagator $D$ of the longitudinal spin fluctuation. 
Detailed analyses of the side-jump contribution $\sigma_{\rm SH}^{\rm side \, jump}$ and 
the skew-scattering contribution $\sigma_{\rm SH}^{\rm skew \, scatt.}$ are presented in the following subsections. 

\subsubsection{Side jump}

The SH conductivity by the side-jump mechanism as diagramatically shown in Fig.~2 
is expressed in terms of the electron Green's function and the propagator of the spin fluctuation as 
\begin{eqnarray}
\sigma^{\rm side \, jump}_{\rm AFM, SH} (\i \Omega_l) 
\!\!\!&=&\!\!\! \frac{1}{\i \Omega_l} \frac{2e^2}{m} \frac{T^2}{VN} \sum_{l, l'} 
\sum_{\vect k, \vect k'} 
\Bigl\{{\cal F}_0 k_x^2 - 2 {\cal F}_1 k_x^2 \bigl( k'_x \bigr)^2 \Bigr\} {\cal F}_2  D_{\vect k'-\vect k}(\i \varepsilon_{l'} -\i \varepsilon_l) \nonumber \\
&\times&\!\!\! G_{\vect k} (\i \varepsilon_l) G_{\vect k}(\i \varepsilon_l + \i \hbar \Omega_l) 
\bigl\{ G_{\vect k'}(\i \varepsilon_{l'}+\i \hbar \Omega_l) - G_{\vect k'}(\i \varepsilon_{l'}) \bigr\}
\label{eq:sigmaSJ0}
\end{eqnarray}
where, $G_{\vect k}(\i \varepsilon_l)=\{\i \varepsilon_l-\varepsilon_{\vect k} - \Sigma_{\vect k} (\i \varepsilon_l)\}^{-1}$ 
is the electron Matsubara Green's function, 
with the fermionic Matsubara frequency $\varepsilon_l = (2l+1) \pi T$. 
The Planck constant $\hbar$ is included explicitly in front of the Matsubara frequency $\Omega_l$.

After carrying out the Matsubara summation, and taking the limit of $\i \Omega_l \rightarrow 0$, one obtains 
\begin{eqnarray}
\sigma^{\rm side \, jump}_{\rm AFM, SH}\!\!\!&=&\!\!\! \frac{2e^2 \hbar}{\pi m V N} \! \int \!\! d\varepsilon d \omega 
\sum_{\vect k, \vect k'} 
\Bigl\{{\cal F}_0 k_x^2 - 2 {\cal F}_1 k_x^2 \bigl( k'_x \bigr)^2 \Bigr\} {\cal F}_2 
 B_{\vect k' -\vect k}(\omega) \Bigl\{ b(\omega) + f \bigl(\varepsilon_{\vect k'} \bigr) \Bigr\}  \nonumber \\
&\times&\!\!\! \Bigl[\partial_\varepsilon f(\varepsilon) 
\bigl| G_{\vect k}^{\rm R}(\varepsilon) \bigr|^2
\Im \bigl\{ G_{\vect k'}^{\rm R}(\varepsilon+\omega) \bigr\} 
+ f(\varepsilon) \Im \bigl\{ G_{\vect k}^{\rm R} (\varepsilon) G_{\vect k}^{\rm R}(\varepsilon) 
\partial_\varepsilon G_{\vect k'}^{\rm R}(\varepsilon+\omega) \bigr\} \Bigr] .
\label{eq:sigmaSJ1}
\end{eqnarray}
$f(\varepsilon)$ and $b(\omega)$ are the Fermi distribution function and the Bose distribution function, respectively. 
$G_{\vect k}^{\rm R,A}(\varepsilon) = G_{\vect k}(\i \varepsilon_l \rightarrow \varepsilon \pm \i \hbar /2 \tau_{\vect k})$ 
are the retarded and advanced Green's function, respectively. 
Here, the self-energy is assumed to be independent of $\varepsilon$, and $\tau_{\vect k}$ is the carrier lifetime. 
$B_{\vect p}(\omega)$ is the spectral function of the $J$ propagator 
given by $B_{\vect p}(\omega)= -\frac{1}{\pi} \Im D_{\vect p}(\i \omega_l \rightarrow \omega +\i 0^+) 
= \frac{1}{\pi} \frac{\omega/\Gamma}{\bigl\{ \delta + A |\vect p - \vect Q|^2 \bigr\}^2 +(\omega/\Gamma)^2}$. 

The first term in the square bracket of Eq.~(\ref{eq:sigmaSJ1}) is proportional to $\partial_\varepsilon f(\varepsilon) \approx - \delta(\varepsilon)$, the so-called Fermi surface term, 
while the second term is proportional to $f(\varepsilon)$, the so-called Fermi sea term. 
In principle, two terms contribute, but it can be shown that the contribution from the second term, the Fermi sea term, is small. 
Thus, we focus on the first contribution. 

We use the following approximations considering the small self-energy $\Sigma_{\vect k} (\varepsilon) = \i \hbar/2 \tau_{\vect k}$: 
$|G^{\rm R}_{\vect k} (\varepsilon)|^2 \approx (2 \pi \tau_{\vect k} /\hbar) \delta(\varepsilon - \varepsilon_{\vect k}) $ and 
$\Im G^{\rm R}_{\vect k} (\varepsilon) \approx -\pi \delta(\varepsilon - \varepsilon_{\vect k}) $. 
Carrying out $\varepsilon$ integral, one finds 
\begin{eqnarray}
\sigma^{\rm side \, jump}_{\rm AFM, SH}= - \frac{2e^2 \hbar}{\pi m V N} \! \int \!\! d \omega 
\sum_{\vect k, \vect k'} 
\Bigl\{{\cal F}_0 k_x^2 - 2 {\cal F}_1 k_x^2 \bigl( k'_x \bigr)^2 \Bigr\} {\cal F}_2 
 B_{\vect k' -\vect k}(\omega) \bigl\{ b(\omega) + f (\omega ) \bigr\}  
\bigl| G_{\vect k}^{\rm R} (0) \bigr|^2
\Im \bigl\{ G_{\vect k'}^{\rm R}(\omega) \bigr\} 
.
\label{eq:sigmaSJ2}
\end{eqnarray}
For the sake of convenience, we change the momentum variable from $\vect k' - \vect k$ to $\vect q$ and 
the notation from $B_{\vect q}(\omega) = \frac{1}{\pi} \frac{\omega/\Gamma}{\bigl( \delta + A |\vect q - \vect Q|^2 \bigr)^2 +(\omega/\Gamma)^2}$
to $B_{\vect q}(\omega) = \frac{1}{\pi} \frac{\omega/\Gamma}{\bigl( \delta + A q^2 \bigr)^2 +(\omega/\Gamma)^2}$, 
where $\vect q$ is the momentum variable measured from magnetic wave vector $\vect Q$. 
Thus, Eq. (\ref{eq:sigmaSJ2}) is re-expressed as 
\begin{eqnarray}
\sigma^{\rm side \, jump}_{\rm AFM, SH} = - \frac{2e^2 \hbar}{\pi m V N} \! \int \!\! d \omega 
\sum_{\vect k, \vect q} 
F_{\vect k,\vect q} (\omega)
%
\bigl| G_{\vect k}^{\rm R} (0) \bigr|^2
\Im \bigl\{ G_{\vect k + \vect Q + \vect q}^{\rm R}(\omega) \bigr\},  
\label{eq:sigmaSJ3}
\end{eqnarray}
with 
$F_{\vect k,\vect q} (\omega) 
= \Bigl\{{\cal F}_0 k_x^2 - 2 {\cal F}_1 k_x^2 \bigl( k_x + Q_x + q_x \bigr)^2 \Bigr\} {\cal F}_2 
B_{\vect q}(\omega) \bigl\{ b(\omega) + f (\omega ) \bigr\} $.

Using $\bigl| G_{\vect k}^{\rm R} (0) \bigr|^2 =  (2 \pi \tau_{\vect k} /\hbar) \delta( \varepsilon_{\vect k})$ 
and $\Im \bigl\{ G_{\vect k + \vect Q + \vect q}^{\rm R} (\omega) \bigr\} = -\pi \delta(\omega - \varepsilon_{\vect k + \vect Q + \vect q})$
explicitly, with $\varepsilon_{\vect k + \vect Q + \vect q} = \frac{\hbar^2}{2m} (\vect k + \vect Q + \vect q)^2 - \varepsilon_{\rm F}$, 
one proceeds to the $\vect k$ integral
to find
\begin{eqnarray}
\sigma^{\rm side \, jump}_{\rm AFM, SH} \!\!\!&=&\!\!\! \frac{1}{N} \frac{e^2 k_{\rm F} }{2 \pi^2  \hbar^2}
\int \!\! d \omega \sum_{\vect q} \int_0^{2 \pi} \!\! d\varphi \int_0^\pi \!\! \sin \theta \, d\theta  \,
\tau_{\vect k} \, F_{\vect k, \vect q} (\omega) 
\nonumber \\
&\times&\!\!\! 
\delta \biggl( \omega - \frac{\hbar^2}{2m} \bigl\{ 2 (\vect k + \vect Q) \cdot \vect q + |\vect Q|^2 + |\vect q|^2 + 2 \cos \theta k_{\rm F} |\vect Q| \bigr\} \biggr) \biggl|
_{|\vect k|=k_{\rm F}}
\, 
\end{eqnarray}
where $\varphi$ is the azimuth angle and $\theta$ is the polar angle of $\vect k$ measured from the $\vect Q$ direction. 
Here, the $\vect k$ dependence is kept in the $\delta$ function, 
but the size of $\vect k$ is constrained as $|\vect k|=k_F = \sqrt{2 m \varepsilon_{\rm F}}/\hbar$. 
By further carrying out the $\theta$ integral, one arrives at 
\begin{eqnarray}
\sigma^{\rm side \, jump}_{\rm AFM, SH} \!\!\!&=&\!\!\! \frac{1}{N} \frac{e^2 m k_{\rm F}}{2 \pi^2 \hbar^4 |\vect Q|}
\int \!\! d \omega \sum_{\vect q} \int_0^{2 \pi} \!\! d\varphi  \,
 \, \tau_{\vect k} F_{\vect k, \vect q} (\omega) 
\Bigl|
%
%
^{|\vect k|=k_{\rm F}}
_{\cos \theta = \frac{m}{\hbar^2 k_{\rm F} |\vect Q|} \Bigl[ \omega - \frac{\hbar^2}{2m} \bigl\{ 2 (\vect k + \vect Q)\cdot \vect q + |\vect Q|^2+ |\vect q|^2 \bigr\} \Bigr]}, 
\end{eqnarray}
where we assumed that $|\vect Q| < 2 k_{\rm F}$. 

Now we consider low temperature $T$ regime, where only small frequency $\omega$ and small momentum $|\vect q|$ contribute to this integral. 
In this case, the constraint 
$\cos \theta = \frac{m}{\hbar^2 k_{\rm F} |\vect Q|} \Bigl[ \omega - \frac{\hbar^2}{2m} \bigl\{ 2 (\vect k + \vect Q) \cdot \vect q + |\vect Q|^2  + |\vect q|^2\bigr\} \Bigr]$ 
can be approximated as 
$\cos \theta \approx -|\vect Q|/2k_{\rm F}$ because contributions coming from $\omega$ and $\vect q$ are expected to induce only small 
and higher-order corrections in $\sigma^{\rm side \, jump}_{\rm AFM, SH}$. 
Together with $|\vect k|=k_{\rm F}$, the condition $\cos \theta \approx -|\vect Q|/2k_{\rm F}$ means 
that the main contribution to $\sigma^{\rm side \, jump}_{\rm SH}$ comes from the momenta that satisfy the nesting condition. 
Further, $q_x$ in ${\cal F}_0 k_x^2 - 2 {\cal F}_1 k_x^2 \bigl( k_x + Q_x + q_x \bigr)^2$ gives 
only small contributions to the spin-Hall conductivity. 
Thus, assuming that the angle dependence of $\tau_{\vect k}$ is weak, the leading term of $\sigma^{\rm side \, jump}_{\rm AFM, SH}$ is summarized as 
\begin{eqnarray}
\sigma^{\rm side \, jump}_{\rm AFM, SH} = 
\frac{e^2 m \tau_{\vect k}}{2 \pi^2 \hbar^4 |\vect Q|} \, A_{\rm AFM}^{\rm side \, jump} \, \tilde I_{\rm AFM} (T,\delta), 
\end{eqnarray}
where 
\begin{eqnarray}
A_{\rm AFM}^{\rm side \, jump} =\int_0^{2 \pi} \!\! d\varphi 
\bigl\{ {\cal F}_0 k_x^2 - 2 {\cal F}_1 k_x^2 (k_x + Q_x)^2 \bigr\} {\cal F}_2
\bigl|
^{|\vect k|=k_{\rm F}}
_{\cos \theta = -|\vect Q|/ 2 k_{\rm F}}, 
\end{eqnarray}
and 
\begin{eqnarray}
\tilde I_{\rm AFM} (T,\delta) \!\!\!&=&\!\!\! \frac{\pi}{N} \sum_{\vect q} \int \!\!  d\omega  \,
B_{\vect q} (\omega) \,\bigl\{ b(\omega)+f(\omega) \bigr\} \nonumber \\
&=&\!\!\!
\frac{a^3}{8 \pi^2} \int \!\! d\omega \!\! \int \!\! d^3 \vect q \, 
\frac{\omega/\Gamma}{\bigl( \delta + A q^2 \bigr)^2 +(\omega/\Gamma)^2}
\,\frac{1}{\sinh (\omega/T)}. 
\label{eq:tildeIAFM}
\end{eqnarray}

\subsubsection{Skew scattering}

Using Matsubara Green's functions for conduction electrons and the spin fluctuation propagator, 
the SH conductivity due to the skew scattering is expressed as 
\begin{eqnarray}
\sigma^{\rm skew \, scat.}_{\rm AFM, SH} (\i \Omega_l) \!\!\!&=&\!\!\! \frac{1}{\i \Omega_l} \frac{e^2 \hbar^2}{m^2} \frac{T^3}{V N^2} \sum_{l, l', l''} 
\sum_{\vect k, \vect k', \vect k''} 
\nonumber \\
&\times&\!\!\! G_{\vect k} (\i \varepsilon_l) G_{\vect k}(\i \varepsilon_l + \i \hbar \Omega_l) 
 G_{\vect k'} (\i \varepsilon_{l'}) G_{\vect k'}(\i \varepsilon_{l'} + \i \hbar \Omega_l) 
\bigl\{ G_{\vect k''}(\i \varepsilon_{l''}+\i \hbar \Omega_l) - G_{\vect k''}(\i \varepsilon_{l''}) \bigr\} \nonumber \\
&\times&\!\!\! \Bigl\{ - D_{\vect k-\vect k''}(\i \varepsilon_l - \i \varepsilon_{l''}) D_{\vect k'-\vect k''}(\i \varepsilon_{l'} - \i \varepsilon_{l''}) 
{\cal F}_3 k_x^2 {k'_y}^2 \bigl( {\cal F}_0 + {\cal F}_1 \vect k \cdot \vect k'' \bigr) \bigl( {\cal F}_0 + {\cal F}_1 \vect k' \cdot \vect k'' \bigr) \nonumber \\
& +&\!\!\! 2 D_{\vect k-\vect k'}(\i \varepsilon_l - \i \varepsilon_{l'}) D_{\vect k''-\vect k'}(\i \varepsilon_{l''} - \i \varepsilon_{l'}) 
{\cal F}_3 k_x^2 {k''_y} {k'_y} \bigl( {\cal F}_0 + {\cal F}_1 \vect k \cdot \vect k' \bigr) \bigl( {\cal F}_0 + {\cal F}_1 \vect k' \cdot \vect k'' \bigr) 
\Bigr\} .
\label{eq:sigmaSS0}
\end{eqnarray}
Here, the last term in Eq.~(1) is not considered because this term is proportional to $(\vect J_n \cdot \vect J_n) \approx const.$. 
The Matsubara summation can be carried out similarly as in the side-jump mechanism. 
Focusing on the Fermi surface terms, one finds 
\begin{eqnarray}
\sigma^{\rm skew \, scat.}_{\rm AFM, SH} \!\!\!&=&\!\!\! \frac{e^2 \hbar^3}{\pi m^2 V N^2} \! \int \!\! d\varepsilon d \omega d \omega' \!\! 
\sum_{\vect k, \vect k', \vect k''} 
\nonumber \\
&\times&\!\!\! \biggl[ 
{\cal F}_3 k_x^2 {k'_y}^2 \bigl( {\cal F}_0 + {\cal F}_1 \vect k \cdot \vect k'' \bigr) \bigl( {\cal F}_0 + {\cal F}_1 \vect k' \cdot \vect k'' \bigr) 
B_{\vect k - \vect k''}(\omega) \Bigl\{ b(\omega) + f \bigl(\varepsilon_{\vect k} \bigr) \Bigr\} 
B_{\vect k' - \vect k''}(\omega') \Bigl\{ b(\omega') + f \bigl(\varepsilon_{\vect k'} \bigr) \Bigr\} \nonumber \\
&\times&\!\!\! \partial_\varepsilon f(\varepsilon) 
\bigl| G_{\vect k}^{\rm R} (\varepsilon + \omega) \bigr|^2
\bigl| G_{\vect k'}^{\rm R} (\varepsilon + \omega') \bigr|^2 
\Im \bigl\{ G_{\vect k''}^{\rm R} (\varepsilon) \bigr\} 
\nonumber \\
&-&\!\!\! 2 
{\cal F}_3 k_x^2 {k''_y} {k'_y} \bigl( {\cal F}_0 + {\cal F}_1 \vect k \cdot \vect k' \bigr) \bigl( {\cal F}_0 + {\cal F}_1 \vect k' \cdot \vect k'' \bigr) 
B_{\vect k - \vect k'}(\omega) \Bigl\{ b(\omega) + f \bigl(\varepsilon_{\vect k} \bigr) \Bigr\} 
B_{\vect k'' - \vect k'}(\omega') \Bigl\{ b(\omega') + f \bigl(\varepsilon_{\vect k''} \bigr) \Bigr\} \nonumber \\
&\times&\!\!\!  \partial_\varepsilon f(\varepsilon) 
\bigl| G_{\vect k}^{\rm R} (\varepsilon + \omega) \bigr|^2
\bigl| G_{\vect k'}^{\rm R} (\varepsilon) \bigr|^2
\Im \bigl\{ G_{\vect k''}^{\rm R} (\varepsilon+\omega') \bigr\} 
\biggr]. 
\label{eq:sigmaSS1}
\end{eqnarray}

Again, we change momentum variables from $\vect k - \vect k''$ to $\vect q$ and from $\vect k' - \vect k''$ to $\vect q'$ and measure them from 
the magnetic wave vector $\vect Q$, and use the new notation of the spin fluctuation propagator $B$. 
Replacing Green's function by $\delta$ functions and carrying out the $\varepsilon$ integral, one finds 
\begin{eqnarray}
\sigma^{\rm skew \, scat.}_{\rm AFM, SH} \!\!\!&=&\!\!\! \frac{4 \pi^2 e^2 \hbar}{m^2 V N^2} \! \int \!\! d \omega d \omega' \!\! 
\sum_{\vect k, \vect q, \vect q'} \delta \bigl(\varepsilon_{\vect k} \bigr) \delta \bigl(\omega - \varepsilon_{\vect k+\vect Q +\vect q} \bigr)
\delta \bigl(\omega' - \varepsilon_{\vect k+\vect Q +\vect q'} \bigr)
\nonumber \\
&\times&\!\!\!
F_{\vect k, \vect q, \vect q'} 
B_{\vect q}(\omega) \bigl\{ b(\omega) + f (\omega) \bigr\} 
B_{\vect q'}(\omega') \bigl\{ b(\omega') + f (\omega') \bigr\} ,
\label{eq:sigmaSS2}
\end{eqnarray}
where 
\begin{eqnarray}
F_{\vect k, \vect q, \vect q'} 
\!\!\!&=&\!\!\!
{\cal F}_3 \bigl(k_x + Q_x +q_x \bigr)^2 \bigl(k_y+Q_y+q_y\bigr)^2 
\bigl\{ {\cal F}_0 + {\cal F}_1 \vect k \cdot (\vect k + \vect Q +\vect q) \bigr\} \bigl\{ {\cal F}_0 + {\cal F}_1 \vect k \cdot (\vect k+\vect Q +\vect q') \bigr\} 
\tau_{\vect k +\vect Q+\vect q} \, \tau_{\vect k +\vect Q+\vect q'}
\nonumber \\
&-&\!\!\! 2 
{\cal F}_3 \bigl(k_x + Q_x +q_x\bigr)^2 \bigl(k_y + Q_y+q_y \bigr) {k_y} 
\bigl\{ {\cal F}_0 + {\cal F}_1 \vect k \cdot (\vect k+\vect Q +\vect q) \bigr\} \bigl\{ {\cal F}_0 + {\cal F}_1 \vect k \cdot (\vect k +\vect Q +\vect q') \bigr\} 
\tau_{\vect k} \, \tau_{\vect k +\vect Q+\vect q} . \nonumber \\
\end{eqnarray}

Carrying out polar and radial parts of the $\vect k$ integral and $\omega'$ integral, one finds 
\begin{eqnarray}
\sigma^{\rm skew \, scat.}_{\rm AFM, SH} \!\!\!&=&\!\!\! \frac{1}{N^2} \frac{e^2}{2 \pi \hbar^3 |\vect Q|}
 \! \int \!\! d \omega d \omega'  
\sum_{\vect q, \vect q'} 
\int_0^{2\pi} \!\! d\varphi 
\delta \bigl( \omega -\varepsilon_{\vect k +\vect Q+\vect q} -\omega' + \varepsilon_{\vect k +\vect Q+\vect q'} \bigr)
F_{\vect k, \vect q, \vect q'} \nonumber \\
\!\!\!&\times&\!\! \!
B_{\vect q}(\omega) \bigl\{ b(\omega) + f (\omega) \bigr\} 
B_{\vect q'}(\omega') \bigl\{ b(\omega') + f (\omega') \bigr\}
\Bigl|^{|\vect k|=k_F}
_{\cos \theta =-|\vect Q|/2 k_{\rm F}}
\nonumber \\
\!\!\!&=&\!\!\! \frac{1}{N^2} 
\frac{e^2}{2 \pi \hbar^3 |\vect Q|} \! \int \!\! d \omega  
\sum_{\vect q, \vect q'} 
\int_0^{2\pi} \!\! d\varphi 
\, F_{\vect k, \vect q, \vect q'} \,
B_{\vect q}(\omega) \bigl\{ b(\omega) + f (\omega) \bigr\}  \nonumber \\
\!\!\!&\times&\!\!\!
B_{\vect q'} (\omega-\varepsilon_{\vect k +\vect Q+\vect q} + \varepsilon_{\vect k +\vect Q+\vect q' }) 
\bigl\{ b(\omega-\varepsilon_{\vect k +\vect Q+\vect q} + \varepsilon_{\vect k +\vect Q+\vect q'}) 
+ f (\omega- \varepsilon_{\vect k +\vect Q+\vect q}+\varepsilon_{\vect k +\vect Q+\vect q'}) \bigr\} 
\Bigl|^{|\vect k|=k_{\rm F}}
_{\cos \theta =-|\vect Q|/2 k_{\rm F}}
\nonumber \\
\end{eqnarray}

As in the side jump case, we consider low temperatures $T$, where only small $\vect q$ and $\vect q'$ contribute to $\sigma^{\rm skew \, scat.}_{\rm AFM, SH}$. 
With the constraints $|\vect k|=k_{\rm F}$ and $\cos \theta =-|\vect Q|/2 k_{\rm F}$, 
one can approximate $\varepsilon_{\vect k +\vect Q + \vect q}$ as $\varepsilon_{\vect k +\vect Q + \vect q} = \frac{\hbar^2}{2m}|\vect k+\vect Q +\vect q|^2 -\varepsilon_{\rm F} \approx \hbar \vect v_{\rm F} \cdot \vect q$, 
where $\vect v_{\rm F}=\hbar (\vect k + \vect Q)/m$ is the Fermi velocity at $\vect k + \vect Q$. 

Now, we re-write $B_{\vect q'} (\omega-\varepsilon_{\vect k +\vect Q+\vect q} + \varepsilon_{\vect k +\vect Q+\vect q' }) $. 
In the critical regime, $|\vect q|^2$ and $\vect \varepsilon_{\vect k +\vect Q+\vect q'}=\hbar \vect v_{\rm F} \cdot \vect q'$ are small. 
Under this assumption, $B_{\vect q'} (\omega-\varepsilon_{\vect k +\vect Q+\vect q} + \varepsilon_{\vect k +\vect Q+\vect q' }) $ is 
re-written as follows: 
\begin{eqnarray}
B_{\vect q'} (\omega-\varepsilon_{\vect k +\vect Q+\vect q} + \varepsilon_{\vect k +\vect Q+\vect q' }) 
\!\!\!&=&\!\!\!
\frac{1}{\pi} \frac{(\omega-\varepsilon_{\vect k +\vect Q+\vect q} + \varepsilon_{\vect k +\vect Q+\vect q' })/\Gamma}
{\bigl(\delta + A q'^2 \bigr)^2 + 
\bigl\{ \bigl(\omega-\varepsilon_{\vect k +\vect Q+\vect q} + \varepsilon_{\vect k +\vect Q+\vect q' } \bigr)/\Gamma \bigr\}^2} \nonumber \\
\!\!\!&\approx&\!\!\!
\frac{1}{\pi} \frac{\bigl( \omega-\varepsilon_{\vect k +\vect Q+\vect q} + \varepsilon_{\vect k +\vect Q+\vect q' } \bigr)/\Gamma}
{\bigl(\delta + A q'^2 \bigr)^2 + 
\bigl\{ \bigl( \varepsilon_{\vect k +\vect Q+\vect q' } \bigr)/\Gamma \bigr\}^2}
\frac{1}
{1+\frac{\bigl\{ \bigl(\omega - \varepsilon_{\vect k + \vect Q +\vect q}\bigr) /\Gamma \bigr\}^2}
{\bigl(\delta + A q'^2 \bigr)^2 + \bigl\{ \bigl(\varepsilon_{\vect k +\vect Q+\vect q'}\bigr)/\Gamma \bigr\}^2}} \nonumber \\
\!\!\!&\approx&\!\!\!
\frac{1}{\pi} \frac{\bigl( \omega-\varepsilon_{\vect k +\vect Q+\vect q} + \varepsilon_{\vect k +\vect Q+\vect q' }\bigr)/\Gamma}
{\bigl(\delta + A q'^2 \bigr)^2 + 
\bigl\{ \bigl(\varepsilon_{\vect k +\vect Q+\vect q' }\bigr)/\Gamma \bigr\}^2}
\frac{ \bigl(\delta + A q'^2 \bigr)^2 \,\Gamma^2 + \varepsilon_{\vect k +\vect Q+\vect q'}^2}
{\bigl(\delta + A q'^2 \bigr)^2 \, \Gamma^2 + \varepsilon_{\vect k +\vect Q+\vect q'}^2 + \bigl( \omega - \varepsilon_{\vect k + \vect Q +\vect q}\bigr)^2} \nonumber \\
\!\!\!&\approx&\!\!\!
\frac{1}{\pi} \frac{\bigl(\varepsilon_{\vect k +\vect Q+\vect q' }\bigr)/\Gamma}
{\bigl(\delta + A q'^2 \bigr)^2 + 
\bigl\{ \bigl(\varepsilon_{\vect k +\vect Q+\vect q' }\bigr)/\Gamma \bigr\}^2}
\, \delta \bigl( \omega - \varepsilon_{\vect k + \vect Q +\vect q}\bigr) \, \pi 
\Gamma \delta 
\end{eqnarray}
In the second line, a small cross term $\varepsilon_{\vect k+\vect Q+\vect q} (\omega -\varepsilon_{\vect k+\vect Q+\vect q'})$
is neglected, and in the last line small terms $(2 \delta A q'^2 + A^2 q'^4) \Gamma^2 + \varepsilon_{\vect k +\vect Q+\vect q'}^2 $, which are proportional to $q'^2$, 
are neglected and $\Gamma \delta$ is assumed to be small. 
Thus, at low temperatures and near the critical regime, 
$B_{\vect q'} (\omega-\varepsilon_{\vect k +\vect Q+\vect q} + \varepsilon_{\vect k +\vect Q+\vect q' })$ is approximated as 
\begin{eqnarray}
B_{\vect q'} (\omega-\varepsilon_{\vect k +\vect Q+\vect q} + \varepsilon_{\vect k +\vect Q+\vect q' }) \approx 
B_{\vect q'} (\varepsilon_{\vect k +\vect Q+\vect q' }) 
\delta \bigl( \omega - \varepsilon_{\vect k + \vect Q +\vect q}\bigr) \, \pi \Gamma \delta. 
\end{eqnarray}
Note that this approximation cannot be used to describe the behavior precisely at the critical point at finite temperature 
because $B_{\vect q'} (\omega-\varepsilon_{\vect k +\vect Q+\vect q} + \varepsilon_{\vect k +\vect Q+\vect q' })$ becomes zero 
as $\delta \rightarrow 0$, 
while $B_{\vect q'} (\omega-\varepsilon_{\vect k +\vect Q+\vect q} + \varepsilon_{\vect k +\vect Q+\vect q' })$ is expected to remain nonzero. 

With this approximation and neglecting small contributions in $F_{\vect k, \vect q, \vect q'}$ from $\vect q$ and $\vect q'$, 
$\vect q$ and $\vect q'$ integrals can be carried out separately. 
As a result, the leading contribution to $\sigma^{\rm skew \, scat.}_{\rm AFM, SH}$ is given by 
\begin{eqnarray}
\sigma^{\rm skew \, scat.}_{\rm AFM, SH} 
\!\!\!&=&\!\!\! \frac{1}{N^2}  
\frac{e^2}{2 \pi \hbar^3 |\vect Q|}
\sum_{\vect q, \vect q'} 
\int_0^{2\pi} \!\! d\varphi 
F_{\vect k, \vect q=0, \vect q'=0} \,
B_{\vect q}(\varepsilon_{\vect k +\vect Q+\vect q}) \bigl\{ b(\varepsilon_{\vect k +\vect Q+\vect q}) + f (\varepsilon_{\vect k +\vect Q+\vect q}) \bigr\}  \nonumber \\
\!\!\!&\times&\!\!\!
B_{\vect q'} (\varepsilon_{\vect k +\vect Q+\vect q' }) 
\bigl\{ b(\varepsilon_{\vect k +\vect Q+\vect q'}) + f (\varepsilon_{\vect k +\vect Q+\vect q'}) \bigr\} 
\Bigl|
^{|\vect k|=k_{\rm F}}
_{\cos \theta =-|\vect Q|/2 k_{\rm F}}
. \nonumber \\
\end{eqnarray}
Considering the constraints $|\vect k|=k_{\rm F}$ and $\cos \theta =-|\vect Q|/2 k_{\rm F}$ and carrying out momentum integrals separately, 
one arrives at 
\begin{eqnarray}
\sigma^{\rm skew \, scat.}_{\rm AFM, SH} 
\!\!\!&=&\!\!\! 
\frac{e^2 \tau_{\vect k}^2 \, \Gamma \, \delta}{2 \pi^2 \hbar^3 |\vect Q|}
A_{\rm AFM}^{\rm skew \, scatt.}\,
I_{\rm AFM}^2 (T,\delta), 
\end{eqnarray}
where $A_{\rm AFM}^{\rm skew \, scatt.}$ and 
$I_{\rm AFM} (T,\delta)$ are given by 
\begin{eqnarray}
A_{\rm AFM}^{\rm skew \, scatt.} \!\!\!&=&\!\!\! \int_0^{2\pi} \!\! d\varphi \Bigl[
{\cal F}_3 \bigl(k_x + Q_x  \bigr)^2 \bigl(k_y+Q_y\bigr)^2 
\bigl\{ {\cal F}_0 + {\cal F}_1 \vect k \cdot (\vect k + \vect Q ) \bigr\} 
\bigl\{ {\cal F}_0 + {\cal F}_1 \vect k \cdot (\vect k+\vect Q ') \bigr\} 
\nonumber \\
&-&\!\!\! 2 
{\cal F}_3 \bigl(k_x + Q_x\bigr)^2 \bigl(k_y + Q_y  \bigr) {k_y} 
\bigl\{ {\cal F}_0 + {\cal F}_1 \vect k \cdot (\vect k+\vect Q ) \bigr\} 
\bigl\{ {\cal F}_0 + {\cal F}_1 \vect k \cdot (\vect k +\vect Q ) \bigr\} \Bigr]
\Bigl|^{|\vect k|=k_{\rm F}}
_{\cos \theta =-|\vect Q|/2 k_{\rm F}}
\end{eqnarray}
and
\begin{eqnarray}
I_{\rm AFM} (T,\delta) \!\!\!&=&\!\!\! 
\frac{\pi}{N} \sum_{\vect q}
B_{\vect q} (\hbar \vect v_{\rm F} \cdot \vect q) \bigl\{ b(\hbar \vect v_{\rm F} \cdot \vect q) + r(\hbar \vect v_{\rm F} \cdot \vect q)\bigr\} \nonumber \\
&=& \frac{a^3}{8 \pi^2}\!\!\! \int d^3 \vect q
\frac{\hbar \vect v_{\rm F} \cdot \vect q/\Gamma}{\bigl( \delta + A q^2\bigr)^2 + \bigl(\hbar \vect v_{\rm F} \cdot \vect q/\Gamma \bigr)^2}
\frac{1}{\sinh \bigl( \hbar \vect v_{\rm F} \cdot \vect q/T \bigr)}, 
\label{eq:IAFM}
\end{eqnarray}
respectively.

\subsubsection{Detail of $\tilde I_{AFM} (T,\delta)$}

In this subsection, we describe the detailed behavior of $\tilde I_{\rm AFM}(T,\delta)$ 
and their analytic forms where these are available. 
These results and other functions are summarized in Supplementary Table~\ref{tab:tableI}. 

Carrying out the azimuth and polar integral of $\vect q$ of Eq.~(\ref{eq:tildeIAFM}), $\tilde I_{\rm AFM}(T,\delta)$ becomes 
\begin{eqnarray}
\tilde I_{\rm AFM}(T,\delta) =
\frac{a^3}{2 \pi^2} \int \!\! d\omega \!\! \int \!\! d q \, 
\frac{q^2 \omega/\Gamma}{\bigl( \delta + A q^2 \bigr)^2 +(\omega/\Gamma)^2}
\,\frac{1}{\sinh (\omega/T)}. 
\end{eqnarray}

(1) {\it Disordered regime or ordered phase away from the finite-temperature phase transition, so that $|\delta| \gg T/\Gamma$}. 

This include $T$ regimes I at $T>T_{\rm N}$ and II at $T<T_{\rm N}$ with nonzero $T_{\rm N}$. 
In this regime, 
the damping term $\omega/\Gamma$ can be neglected compared with $\delta$, so that
\begin{eqnarray}
\tilde I_{\rm AFM}(T,\delta) \approx
\frac{a^3}{2 \pi^2} \int_{-\infty}^\infty \!\! d\omega \!\! \int_0^\infty \!\! d q \, 
\frac{q^2 \omega/\Gamma}{\bigl( \delta + A q^2 \bigr)^2}
\,\frac{1}{\sinh (\omega/T)}.  
\end{eqnarray}

Making use of $\int_0^\infty dx \frac{x}{\sinh \,x} = \frac{\pi^2}{4}$, one finds 
\begin{eqnarray}
\tilde I_{\rm AFM}(T,\delta) \approx
\frac{a^3 T^2}{4\Gamma} \int_0^\infty \!\! d q \, 
\frac{q^2}{\bigl( \delta + A q^2 \bigr)^2}
= \frac{a^3 T^2}{4\Gamma} \frac{1}{2 A \sqrt{A \delta}} \arctan \frac{q}{\sqrt{A \delta}}\biggr]_0^\infty = \frac{\pi a^3 T^2}{16 \Gamma A \sqrt{A \delta}}. 
\end{eqnarray}

(2) {\it Near the finite-temperature phase transition, so that $|\delta| \ll T/\Gamma$}. 

In this regime, $\delta$ can be neglected in the frequency regime at $\omega/\Gamma \gg \delta$. 
Considering small but not too small temperature $T$, we approximate $\frac{\omega}{\sinh (\omega/T)}$ as $\frac{\omega}{\sinh (\omega/T)} \approx T$. 
Then, we introduce the upper limit of the frequency integral as $T$, divide the frequency range at $\omega = \Gamma \delta$, and set $\delta = 0$ above $\omega = \Gamma \delta$. 

\begin{eqnarray}
\tilde I_{\rm AFM} (T,\delta) \!\!&\approx&\!\! \frac{a^3 T}{\pi^2 \Gamma} \int_0^\infty dq \, q^2 \biggl[ \int_0^{\delta \Gamma} d \omega \frac{1}{(\delta + A q^2)^2 + (\omega/\Gamma)^2} 
+ \int_{\delta \Gamma}^T d \omega \frac{1}{(A q^2)^2 + (\omega/\Gamma)^2} \biggr] \nonumber \\
&=&\!\! \frac{a^3 T}{\pi^2} \int_0^\infty dq \, q^2 \biggl[ \frac{1}{\delta + A q^2} \arctan \frac{\delta}{\delta + A q^2} + \frac{1}{A q^2} \biggl( \arctan \frac{T}{A q^2 \Gamma} -\arctan\frac{\delta}{Aq^2} \biggr)
\biggr] \nonumber \\
&=&\!\! \frac{a^3 T}{\pi^2} \int_0^\infty dq \, q^2 \frac{1}{\delta + A q^2} \arctan \frac{\delta}{\delta + A q^2} + \frac{T}{\sqrt{2 A^3}} \biggl( \sqrt{\frac{T}{\Gamma}} - \sqrt{\delta}\biggr). 
\end{eqnarray}
In the last line, we made use of 
\begin{eqnarray}
\int_0^\infty \!\! dx \arctan \frac{a}{x^2} = x \arctan \frac{a}{x^2}\biggr]_0^\infty +\int_0^\infty \!\! dx \frac{2a x^2}{x^4+a^2} = \pi \sqrt{\frac{a}{2}}. 
\end{eqnarray}

Since we are focusing at small $|\delta|$, approximating $\arctan \frac{\delta}{\delta + A q^2}$ as $\arctan \frac{\delta}{\delta + A q^2} \approx \frac{\delta}{\delta + A q^2}$ 
and $\int_0^\infty dq \frac{a^2 }{(\delta + A q^2)^2} = \frac{\pi}{4 \sqrt{A^3 \delta}}$, we arrive at 
\begin{eqnarray}
\tilde I_{\rm AFM}(T,\delta) \approx \frac{a^3 T}{\pi \sqrt{2A^3}}\biggl[ \sqrt{\frac{T}{\Gamma}} - \biggl(1-\frac{1}{2 \sqrt{2}}\biggr) \sqrt{\delta} \biggr]. 
\label{eq:I_critical}
\end{eqnarray}

(3) {\it Quantum critical regime}. 

This corresponds to $T$ regime III. 
In this regime, 
one expects $\delta \propto T^{3/2} \ll T$. Therefore, the first term of Eq.~(\ref{eq:I_critical}) becomes dominant. 
Therefore, 
\begin{eqnarray}
\tilde I_{\rm AFM}(T,\delta) \approx \frac{a^3 T}{\pi \sqrt{2A^3}} \sqrt{\frac{T}{\Gamma}}. 
\label{eq:I_quantumcritical}
\end{eqnarray}

Similar temperature dependence was suggested to appear in the electrical resistivity in Ref.~\cite{SUeda1977}. 

\begin{table*}
\caption{$T$ dependence of $\delta$ and $T$-$\delta$ dependence of $\tilde I_{\rm AFM} (T,\delta)$, $I_{\rm AFM}(T,\delta)$, and $I_{\rm FM}(T,\delta)$. 
$const.$ is a dimensionless constant, which depends on how to introduce the frequency cutoff. 
In the analysis highlighted in the supporting information, $const. = (1-\frac{1}{2\sqrt{2}})$. 
Ref.~\cite{SWoelfle2021} reports $const.=3$. }
\begin{tabular}{cc|cccc}
\hline
&&&&\\[-0.8em]
\multicolumn{2}{c|}{$T$ regime} & $\delta$ & $\tilde I_{\rm AFM} (T, \delta)$ & $I_{\rm AFM}(T,\delta)$ & $I_{\rm FM}(T,\delta)$\\
\hline \hline
&&&&\\[-0.8em]
I & $T>T_{\rm N,C}$ & $\sim T-T_{\rm N,C }$ & 
$\approx \frac{\pi}{16} \frac{a^3 T^2}{\Gamma \sqrt{A^3 \delta}}$ &
$\approx \frac{1}{8\pi} \bigl(\frac{aT}{\hbar v_{\rm F}} \bigr)^3 \frac{1}{\delta}$ &
$\approx \frac{1}{8\pi} \bigl(\frac{aT}{\hbar v_{\rm F}} \bigr)^3 \frac{1}{\delta}$ \\
&&&&\\[-0.8em]
--- & $T \sim T_{\rm N,C}$ & $\sim |T-T_{\rm N,C}|$ & 
$ \hspace{1mm} \approx \frac{a^3 T}{\pi \sqrt{2 A^3}} \Bigl[\sqrt{\frac{T}{\Gamma}} - const. \sqrt{\delta} \Bigr]$ &
$ \hspace{1mm} \approx \frac{a^3 T}{8\pi \hbar v_{\rm F} A} \ln \bigl|\frac{A}{\delta} \bigl(\frac{T}{\hbar v_{\rm F}} \bigr)^2 \bigr|$ &
$  \approx \frac{a^3 T}{8\pi \Gamma q_{\rm c}} \frac{1}{A \sqrt{A\delta}}$
\\
&&&&\\[-0.8em]
II & $T<T_{\rm N,C}$ & $ \hspace{1mm} \propto M^2(T) \sim T_{\rm N,C}-T$ & 
$\approx \frac{\pi}{16} \frac{a^3 T^2}{\Gamma \sqrt{A^3 \delta}}$&
$\approx \frac{1}{8\pi} \bigl(\frac{aT}{\hbar v_{\rm F}} \bigr)^3 \frac{1}{\delta}$&
$\approx \frac{1}{8\pi} \bigl(\frac{aT}{\hbar v_{\rm F}} \bigr)^3 \frac{1}{\delta}$\\
[-0.5em]
\multicolumn{2}{l|}{\hspace{-0.5em} \line(1,0){1em} \, \line(1,0){1em} \, \line(1,0){1em} \, \line(1,0){1em} \, 
\line(1,0){1em} \, \line(1,0){1em} \, \line(1,0){1em}} & 
\multicolumn{4}{l}{\line(1,0){1em} \, \line(1,0){1em} \, \line(1,0){1em} \, \line(1,0){1em} \, \line(1,0){1em} \, \line(1,0){1em} \, 
\line(1,0){1em} \, \line(1,0){1em} \, \line(1,0){1em} \, \line(1,0){1em} \, \line(1,0){1em} \,
\line(1,0){1em} \, \line(1,0){1em} \, \line(1,0){1em} \, \line(1,0){1em} \, \line(1,0){1em} \,
\line(1,0){1em} \, \line(1,0){1em} \, \line(1,0){1em} \, \line(1,0){1em} \, \line(1,0){1em} \,
\line(1,0){1em} \, \line(1,0){1em} \, \line(1,0){1em} \, \line(1,0){1em} \, \line(1,0){1em} \,
\line(1,0){1em}     }  
\\
III & $T \sim 0$ ($T_{\rm N,C} \rightarrow 0$) & $\sim T^{3/2}$* & 
$\approx \frac{a^3 T}{\pi \sqrt{2 A^3}} \sqrt{\frac{T}{\Gamma}}$ &
$\approx \frac{1}{8\pi} \bigl(\frac{aT}{\hbar v_{\rm F}} \bigr)^3 \frac{1}{\delta}$ &
$\approx \frac{1}{8\pi} \bigl(\frac{aT}{\hbar v_{\rm F}} \bigr)^3 \frac{1}{\delta}$ \\
\hline
\end{tabular}
\begin{tabular}{rl}
* & For FM with no disorder effects, $\delta \propto T^{4/3}$. 
\end{tabular}
\label{tab:tableI}
\end{table*}

\subsubsection{Detail of $I_{AFM} (T,\delta)$}

In this subsection, we describe the detailed behavior of $I_{\rm AFM}(T,\delta)$. 
We first consider finite but relatively low temperature regime, where the $A$ term is relevant and 
the linear approximation $\varepsilon_{\vect k'}\approx \hbar \vect v_{\rm F} \cdot \vect q$ can be applied.  
We further approximate $ 1/\sinh \bigl( \hbar \vect v_{\rm F} \cdot \vect q/T \bigr) \approx T/ \hbar \vect v_{\rm F} \cdot  \vect q$. 
Considering a three dimensional system, the $\vect q$ integral in Eq.~(\ref{eq:IAFM}) is evaluated as 
\begin{eqnarray}
I_{\rm AFM}(T,\delta)
&\approx&\!\!\! 
\frac{a^3}{(2 \pi)^3}\!\!\! \int d^3 \vect q
\frac{T/\Gamma}{\bigl( \delta+A q^2\bigr)^2 + \bigl( \hbar \vect v_{\rm F} \cdot \vect q/\Gamma \bigr)^2} 
=
\frac{a^3}{(2 \pi)^3} \int_0^{T/\hbar v_{\rm F}} \!\!\!\! dq \int_{0}^\pi  \!\!\! d\theta 
\frac{ 2 \pi q^2 T \sin \theta /\Gamma}{\bigl( \delta+A q^2\bigr)^2 + \bigl(\hbar v_{\rm F} q \cos \theta /\Gamma \bigr)^2} \nonumber \\
&=&\!\!\! 
\frac{a^3 T}{4 \pi^2 \Gamma}\int_0^{T/\hbar v_{\rm F}} \!\!\!\! dq \int_{-1}^1 \!\!\! dx 
\frac{q^2}{\bigl( \delta+A q^2\bigr)^2 + \bigl(\hbar v_{\rm F} q x /\Gamma \bigr)^2} . 
\label{eq:IAFM2}
\end{eqnarray}
The range of $x$ integral can be extended to $(-\infty,\infty)$, allowing the analytic integral over $q$. 
This leads to the following form: 
\begin{eqnarray}
I_{\rm AFM}(T,\delta)&\approx&\!\!\! 
\frac{a^3 T}{8 \pi \hbar v_F A} \ln \biggl| \frac{A}{\delta} \biggl( \frac{T}{\hbar v_{\rm F}} \biggr)^2 +1 \biggr| . 
\label{eq:qint}
\end{eqnarray}

(1) {\it Disordered regime or ordered phase away from the finite-temperature phase transition, so that $|\delta| \gg A (T/\hbar v_{F})^2$}. 

This include $T$ regimes I at $T>T_{\rm N}$ and II at $T<T_{\rm N}$ with nonzero $T_{\rm N}$. 
In this regime, $\ln$ can be expanded, so that 
\begin{eqnarray}
I_{\rm AFM}(T,\delta) \approx \frac{1}{8 \pi} \biggl( \frac{a T}{\hbar v_{\rm F}} \biggr)^3 \frac{1}{\delta}. 
\label{eq:tildeI_disorder}
\end{eqnarray}

(2) {\it Near the finite-temperature phase transition, so that $|\delta| \ll A (T/\hbar v_{F})^2$}. 

In this regime, opposite to the case (1), $1$ inside the logarithmic function can be neglected, so tat
\begin{eqnarray}
I_{\rm AFM}(T,\delta)&\approx&\!\!\! 
\frac{a^3 T}{8 \pi \hbar v_{\rm F} A} \ln \biggl| \frac{A}{\delta} \biggl( \frac{T}{\hbar v_{\rm F}} \biggr)^2 \biggr| . 
\label{eq:tildeI_transition}
\end{eqnarray}
Thus, compared with Eq.~(\ref{eq:tildeI_disorder}), the divergence with reducing $\delta$ is weaker.

(3) {\it Quantum critical regime}. 

Now, we consider quantum critical regime at low temperatures, the $T$ regime III. 
In this case, 
the $A q^2$ term is irrelevant compared with $\delta$ and $\hbar \vect v_{\rm F} \cdot \vect q/\Gamma$ terms. 
Considering a three dimensional system, the $\vect q$ integral is evaluated as 
\begin{eqnarray}
I_{\rm AFM}(T,\delta)
&\approx&\!\!\! 
\frac{a^3 }{(2 \pi)^3}\!\!\! \int d^3 \vect q
\frac{T/\Gamma}{\delta^2 + \bigl( \hbar \vect v_{\rm F} \cdot \vect q/\Gamma \bigr)^2} 
=
\frac{a^3 T}{4 \pi^2 \Gamma} \int_0^{T/\hbar v_{\rm F}} \!\!\!\! dq \int_{0}^\pi  \!\!\! d\theta 
\frac{q^2 \sin \theta}{\delta^2 + \bigl(\hbar v_{\rm F} q \cos \theta /\Gamma \bigr)^2} \nonumber \\
&\approx&\!\!\! 
\frac{a^3 T}{4 \pi^2 \Gamma}\int_0^{T/\hbar v_{\rm F}} \!\!\!\! dq \int_{-\infty}^\infty \!\! dx 
\frac{q^2}{\delta^2 + \bigl(\hbar v_{\rm F} q x /\Gamma \bigr)^2} 
= \frac{a^3 T}{4 \pi} \int_0^{T/\hbar v_{\rm F}} \!\!\! dq \,  \frac{q}{h v_{\rm F} \delta} 
= \frac{1}{8 \pi} \biggl( \frac{a T}{h v_{\rm F}} \biggr)^3 \frac{1}{\delta}. 
\label{eq:tildeI_QC}
\end{eqnarray}
In the second line, the integral range of $x=\sin \theta$ is extended from $(-1,1)$ to $(-\infty, \infty)$. 
The final expression is identical to Eq.~(\ref{eq:tildeI_disorder}).

\subsection{Spin-Hall conductivity of ferromagnets}
\label{subsection:FM}
Here, we consider the ferromagnetic (FM) spin fluctuation~\cite{SOkamoto2019}. 
Considering the periodic lattice system as discussed in the previous sections, the model Hamiltonian is given by Eq.~(1) in the main text. 
The FM spin fluctuation is characterized by the propagator $D$, whose
spectral function $B_{\vect q}(\omega)$ is given by 
$B_{\vect q}(\omega)= -\frac{1}{\pi} \Im D_{\vect q}(\i \omega_l \rightarrow \omega +\i 0^+)
= \frac{1}{\pi} \frac{\omega/\Gamma_{\vect q}}{( \delta + A q^2 )^2 +(\omega/\Gamma_{\vect q})^2}$, 
where the damping term $\Gamma_{\vect q}$ for clean systems is given by $\Gamma q$, with $\Gamma$ is a constant with the dimension of length. 
For systems with impurity scattering or disorder, this damping term has a momentum cutoff at $q=q_{\rm c}$, 
so that $\Gamma_{\vect q}$ is given by $\Gamma_{\vect q}=\Gamma q$ for $q \ge q_{\rm c}$ and $\Gamma_{\vect q}=\Gamma q_{\rm c}$ for $q \le q_{\rm c}$. 

\subsubsection{Side jump}
The SH conductivity by the side-jump-type mechanism as diagramatically shown in Fig.~2 
is expressed in terms of the electron Green's function and the propagator of the spin fluctuation. 
The formal derivation is identical to that of the side-jump due to the AFM spin fluctuation up to Eq.~(\ref{eq:sigmaSJ2}). 

By carrying out $\omega$ integral, one finds

\begin{eqnarray}
\hspace{-1em}\sigma^{\rm side \, jump}_{\rm FM, SH} 
\!\!\!&\approx&\!\!\! \frac{2 e^2 \pi }{m V N} 
\sum_{\vect k, \vect k'} \tau_{\vect k} \delta(\varepsilon_{\vect k})
\Bigl\{{\cal F}_0 k_x^2 - 2 {\cal F}_1 k_x^2 (k'_x)^2 \Bigr\} {\cal F}_2 
B_{\vect k' - \vect k} \bigl( \varepsilon_{\vect k'} \bigr) 
\Bigl\{ b\bigl( \varepsilon_{\vect k'} \bigr) + f \bigl( \varepsilon_{\vect k'} \bigr)\Bigr\} .
\label{eq:sigmaFMSJ0}
\end{eqnarray}
Changing the momentum variable from $\vect k'$ to $\vect k + \vect q$, and approximating the low-energy dispersion as
$\delta (\varepsilon_{\vect k}) \varepsilon_{\vect k + \vect q} \approx \delta (\varepsilon_{\vect k}) \hbar \vect v_{\rm F} \cdot \vect q$,
one arrives at
\begin{eqnarray}
\hspace{-1em}\sigma^{\rm side \, jump}_{\rm FM, SH} 
\!\!\!&\approx&\!\!\! \frac{2 e^2 \pi }{m V N} 
\sum_{\vect k, \vect q} \tau_{\vect k} \delta(\varepsilon_{\vect k})
\Bigl\{{\cal F}_0 k_x^2 - 2 {\cal F}_1 k_x^2 (k_x+q_x)^2 \Bigr\} {\cal F}_2 
B_{\vect q} \bigl( \hbar \vect v_{\rm F} \cdot \vect q \bigr) 
\Bigl\{ b\bigl( \hbar \vect v_{\rm F} \cdot \vect q\bigr) + f \bigl( \hbar \vect v_{\rm F} \cdot \vect q \bigr)\Bigr\} .
\label{eq:sigmaFMSJ1}
\end{eqnarray}
By neglecting small corrections coming from $q_x^2$, the $\vect q$ integral is summarized into the following function, 
\begin{eqnarray}
I_{\rm FM} (T,\delta) \!\!\!&\equiv&\!\!\!
\frac{\pi}{N} \sum_{\vect q} 
B_{\vect q} \bigl( \hbar \vect v_{\rm F} \cdot \vect q \bigr) 
\Bigl\{ b\bigl( \hbar \vect v_{\rm F} \cdot \vect q\bigr) + f \bigl( \hbar \vect v_{\rm F} \cdot \vect q \bigr)\Bigr\}  \nonumber \\
&=&\!\!\!
\frac{a^3}{8 \pi^2}\!\! \int \!\! d^3 {\vect q} \,
\frac{\hbar {\vect v}_{\rm F} \cdot \vect q/\Gamma_{\vect q}}{(\delta + A q^2)^2+(\hbar {\vect v}_{\rm F} \cdot \vect q/\Gamma_{\vect q})}
\frac{1}{\sinh(\hbar {\vect v}_{\rm F} \cdot \vect q/T )} 
\label{eq:IFM}
\end{eqnarray}
with 
\begin{eqnarray}
\Gamma_{\vect q} = \left\{
\begin{array}{ll}
\Gamma q_{\rm c} & (q \le q_{\rm c}) \\
\Gamma q  & (q \ge q_{\rm c})
\end{array}
\right. .
\end{eqnarray}
Combining Eqs.~(\ref{eq:sigmaFMSJ1}) and (\ref{eq:IFM}), one arrives at 
\begin{eqnarray}
\hspace{-1em}\sigma^{\rm side \, jump}_{\rm FM, SH} 
= \frac{2 e^2 \tau_{\vect k}}{m} 
A_{\rm FM}^{\rm side \, jump}
I_{\rm FM}(T,\delta), 
\label{eq:sigmaFMSJ2}
\end{eqnarray}
where
\begin{eqnarray}
A_{\rm FM}^{\rm side\,jump}=\frac{1}{(2\pi)^3}\int \!\!\! d^3 {\vect k} \,
\delta(\varepsilon_{\vect k})
\Bigl\{{\cal F}_0 k_x^2 - 2 {\cal F}_1 k_x^4 \Bigr\} {\cal F}_2 .
\end{eqnarray}
$\tau_{\vect k}$ is the carrier lifetime on the Fermi surface. 
Its momentum dependence is assumed to be independent of $\vect k$, and $\tau_{\vect k}$ is moved outside the $\vect k$ integral. 

\subsubsection{Skew scattering} 
The SH conductivity by the skew-scattering-type mechanism as diagramatically shown in Fig.~3.  
As in the side-jump mechanism, the formal derivation is identical to that of the skew-scattering by the AFM spin fluctuation up to 
Eq.~(\ref{eq:sigmaSS1}). 
By carrying out $\omega$, $\omega'$, and $\varepsilon$ integrals, 
one finds 
\begin{eqnarray}
\sigma^{\rm skew \, scat.}_{\rm FM, SH} \!\!\!&\approx&\!\!\! \frac{4 e^2 \hbar \pi^2 }{m^2 V N^2} \!\! \sum_{\vect k, \vect k', \vect k''} 
\delta(\varepsilon_{\vect k''}) 
B_{\vect k - \vect k''}(\varepsilon_{\vect k}) \Bigl\{ b(\varepsilon_{\vect k}) + f \bigl(\varepsilon_{\vect k} \bigr) \Bigr\} 
B_{\vect k' - \vect k''}(\varepsilon_{\vect k'}) \Bigl\{ b(\varepsilon_{\vect k'}) + f \bigl(\varepsilon_{\vect k'} \bigr) \Bigr\} \nonumber \\
&\times&\!\!\! \bigl\{ 
{\tau_{\vect k} \tau_{\vect k'}
\cal F}_3 k_x^2 {k'_y}^2 \bigl({\cal F}_0 + {\cal F}_1 \vect k \cdot \vect k''\bigr) \bigl ({\cal F}_0 + {\cal F}_1 \vect k' \cdot \vect k'' \bigr) 
-2 \tau_{\vect k} \tau_{\vect k''} 
{\cal F}_3 k_x^2 {k''_y} {k'_y} \bigl( {\cal F}_0 + {\cal F}_1 \vect k \cdot \vect k'' \bigr) \bigl({\cal F}_0 + {\cal F}_1 \vect k' \cdot \vect k'' \bigr) 
\bigr\}. \nonumber \\
\label{eq:sigmaFMSS1}
\end{eqnarray}

By neglecting small corrections coming from $\vect k - \vect k''$ and $\vect k' - \vect k''$, 
one arrives at 
\begin{eqnarray}
\sigma^{\rm skew \, scat.}_{\rm FM, SH} = \frac{4 e^2 \hbar \tau_{\vect k}^2}{m^2} 
A_{\rm FM}^{\rm skew \, scatt.} I_{\rm FM}^2(T,\delta), 
\end{eqnarray}
where 
\begin{eqnarray}
A_{\rm FM}^{\rm skew \, scatt.} \!\!\!&=&\!\!\! 
-\frac{1}{(2 \pi)^3} \!\!\! \int d^3 {\vect k} \, \delta(\varepsilon_{\vect k})  
{\cal F}_3 k_x^2 {k_y}^2 \bigl({\cal F}_0 + {\cal F}_1 |\vect k|^2 \bigr)^2 . 
\end{eqnarray}
As in the side-jump contribution, we neglect the momentum dependence of $\tau_{\vect k}$ and move it outside the momentum integral. 

\subsubsection{Detail of $I_{FM}(T,\delta)$}
Here, we describe the detail of $I_{\rm FM}(T,\delta)$ defined by Eq.~(\ref{eq:IFM}). 
As in $I_{\rm AFM}(T,\delta)$, we consider finite but relatively low temperature regime, where the $A$ term is relevant, 
and $ 1/\sinh \bigl( \hbar \vect v_{\rm F} \cdot \vect q/T \bigr)$ is approximated by $T/ \hbar \vect v_{\rm F} \cdot  \vect q$. 
Considering a three dimensional system, the $\vect q$ integral in Eq.~(\ref{eq:IFM}) is evaluated as 
\begin{eqnarray}
I_{\rm FM}(T,\delta)
&\approx&\!\!\! 
\frac{a^3}{(2 \pi)^3}\!\!\! \int d^3 \vect q
\frac{T/\Gamma_{\vect q}}{\bigl( \delta+A q^2\bigr)^2 + \bigl( \hbar \vect v_{\rm F} \cdot \vect q/\Gamma_{\vect q} \bigr)^2} 
=
\frac{a^3}{(2 \pi)^3} \int_0^{T/\hbar v_{\rm F}} \!\!\!\! dq \int_{0}^\pi  \!\!\! d\theta 
\frac{ 2 \pi q^2 T \sin \theta /\Gamma_{\vect q}}{\bigl( \delta+A q^2\bigr)^2 + \bigl(\hbar v_{\rm F} q \cos \theta /\Gamma_{\vect q} \bigr)^2} \nonumber \\
&=&\!\!\! 
\frac{a^3 T}{4 \pi^2 \Gamma_{\vect q}}\int_0^{T/\hbar v_{\rm F}} \!\!\!\! dq \int_{-1}^1 \!\!\! dx 
\frac{q^2}{\bigl( \delta+A q^2\bigr)^2 + \bigl(\hbar v_{\rm F} q x /\Gamma_{\vect q} \bigr)^2} 
=
\frac{a^3 T}{4 \pi^2 \Gamma_{\vect q}}\int_0^{T/\hbar v_{\rm F}} \!\!\!\! dq 
\frac{2q}{\delta+A q^2} \arctan \frac{\hbar v_{\rm F} q/\Gamma_{\vect q}}{\delta + A q^2}. \nonumber \\
\label{eq:IFM2}
\end{eqnarray}

(1) {\it Disordered regime or ordered phase away from the finite-temperature phase transition, so that $|\delta| \gg A (T/\hbar v_{F})^2$}. 

The calculation is similar to the AFM case. 
Extending the $x$ integral from $(-1,1)$ to $(-\infty,\infty)$ in Eq.~(\ref{eq:IFM2}) or approximating $\arctan$ by $\frac{\pi}{2}$, one finds
\begin{eqnarray}
I_{\rm FM}(T,\delta)&\approx&\!\!\! 
\frac{a^3 T}{8 \pi \hbar v_{\rm F} A} \ln \biggl| \frac{A}{\delta} \biggl( \frac{T}{\hbar v_{\rm F}} \biggr)^2 +1 \biggr| . 
\end{eqnarray}
In the limit of $|\delta| \gg A (T/\hbar v_{\rm F})^2$, this is approximated as 
\begin{eqnarray}
I_{\rm FM}(T,\delta) \approx \frac{1}{8 \pi} \biggl( \frac{a T}{\hbar v_{\rm F}} \biggr)^3 \frac{1}{\delta}. 
\end{eqnarray}

(2) {\it Near the finite-temperature phase transition, so that $|\delta| \ll A (T/\hbar v_{F})^2$}. 

Considering the case where $\hbar v_{\rm F} q/\Gamma_{\vect q} \approx \hbar v_{\rm F}/\Gamma \propto T^0 \ll A q^2 \propto T^2$, 
the argument of $\arctan$ is expanded, and the momentum integral is separated into two regions, $q \le q_{\rm c}$ and $q \ge q_{\rm c}$. 
\begin{eqnarray}
I_{\rm FM}(T,\delta)&\approx&\!\!\! 
\frac{a^3 T}{(2 \pi)^2 \hbar v_{\rm F}} \Biggl\{ 
\int_0^{q_c} \!\!\!\! dq \frac{2q^2 \hbar v_{\rm F}/\Gamma q_{\rm c}}{(\delta + A q^2)^2}  
+ 
\int_{q_{\rm c}}^{T/\hbar v_{\rm F}} \!\!\!\! dq \frac{2q \hbar v_{\rm F}/\Gamma}{(\delta + A q^2)^2} \Biggr\} \nonumber \\
&=&\!\!\! \frac{a^3 T}{(2 \pi)^2 \Gamma } \Biggl\{ 
\frac{1}{q_{\rm c} A \sqrt{A\delta}}\arctan \sqrt{\frac{A}{\delta}}q_{\rm c} - \frac{1}{A \{\delta + A(T/\hbar v_{\rm F})^2\}} 
\Biggr\} 
\end{eqnarray}
Near the phase transition (Curie temperature $T_{\rm C}$), $\delta$ behaves as $\delta \propto |T-T_{\rm C}|$, and 
with $T \rightarrow T_{\rm C}$, $\delta \rightarrow 0$. 
Therefore, the critical behavior of $I_{\rm FM}(T,\delta)$ is given by approximating $\arctan \sqrt{\frac{A}{\delta}}q_{\rm c} \approx \frac{\pi}{2}$, leading to
\begin{eqnarray}
I_{\rm FM}(T,\delta)&\approx&\!\!\! 
\frac{a^3 T}{8 \pi \Gamma q_{\rm c}} 
\frac{1}{A \sqrt{A\delta}}. 
\end{eqnarray}

(3) {\it Quantum critical regime}. 

In this $T$ regime III, the momentum integral is dominated by $q \alt T/ \hbar v_{\rm F}$. 
With the impurity scattering or disorder, the damping term remains finite at small $q$ as $\Gamma_{\vect q} =\Gamma q_{\rm c}$. 
This self-consistently determine that $\delta$ behaves as $\delta \propto T^{3/2}$. 
When disorder effects are absent, $\delta \propto T^{4/3}$ and $q_{\rm c} =0$. 
In either case, the argument of $\arctan$ $\frac{\hbar v_{\rm F} q/\Gamma_{\vect q}}{\delta + A q^2}$ diverges as $T \rightarrow 0$, allowing approximating 
$\arctan \frac{\hbar v_{\rm F} q/\Gamma_{\vect q}}{\delta + A q^2} \approx \frac{\pi}{2}$ as in the disordered case (1). 
Thus, the form of $I_{\rm FM}(T,\delta)$ in this case is identical to that in the disordered case (1) and $I_{\rm AFM}(T,\delta)$ in the quantum critical regime, 
\begin{eqnarray}
I_{\rm FM}(T,\delta)
\approx \frac{1}{8 \pi} \biggl( \frac{a T}{h v_{\rm F}} \biggr)^3 \frac{1}{\delta}. 
\end{eqnarray}
For a disordered case, $\delta \propto T^{3/2}$ leads to $I_{\rm FM} (T,\delta) \propto T^{3/2}$, while for an undisordered case, 
$\delta \propto T^{4/3}$ leads to $I_{\rm FM} (T,\delta) \propto T^{5/3}$. 

The results of $I_{\rm FM} (T,\delta)$ are also summarized in Supplementary Table~\ref{tab:tableI}. 

In Ref.~\cite{SOkamoto2019}, the low-temperature behavior of $I_{\rm FM}(T,\delta)$ with disorder or impurity scattering was evaluated 
considering the limit of $T \ll \Gamma q_{\rm c} \delta$, 
corresponding to approximating $\arctan \frac{\hbar v_{\rm F} q/\Gamma_{\vect q}}{\delta + A q^2}$ in Eq.~(\ref{eq:IFM2})
by $\frac{\hbar v_{\rm F} q/\Gamma_{\vect q}}{\delta + A q^2}$. 
This led to $I_{\rm FM}(T,\delta) \propto T^4/\delta^2$ and, thus, the different power law of $I_{\rm FM}(T,\delta) \propto T$ with $\delta \propto T^{3/2}$. 
However, because the temperature dependence of $\delta$ is $\delta \propto T^{3/2}$ in disordered system, 
$T \ll \Gamma q_{\rm c} \delta$ does not correspond to the proper low temperature limit. 

\subsection{Carrier lifetime by the spin fluctuation} 

\begin{figure}
\begin{center}
\includegraphics[width=0.2\columnwidth, clip]{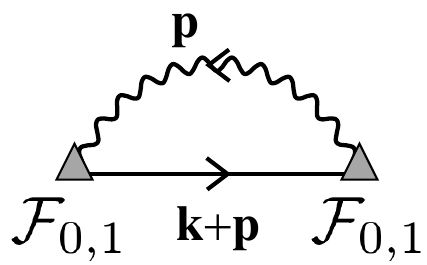}
\caption{Diagrammatic representation for the electron self-energy. 
}
\label{fig:selfenergy}
\end{center}
\end{figure}

\subsubsection{AFM spin fluctuation}
Here, we consider the electron self-energy $\Sigma_{\vect k} (\i \varepsilon_l)$ due to the coupling with the antiferromagnetic spin fluctuation. 
The lowest order self-energy is given by (see Supplementary Figure~\ref{fig:selfenergy} for the diagramatic representation) 
\begin{eqnarray}
\Sigma_{\vect k}(\i \varepsilon_l) = \frac{T}{N} \sum_{\vect p, \i \omega_{l}} 
\Bigl\{ {\cal F}_0 + 2 {\cal F}_1 \bigl( k^2 + \vect k \cdot \vect p \bigr) \Bigr\}^2
G_{\vect k + \vect p}(\i \varepsilon_l + \i \omega_{l}) D_{\vect p} (\i \omega_{l}). 
\end{eqnarray} 
After carrying out the Matsubara summation and the analytic continuation $\i \varepsilon_l \rightarrow \varepsilon + \i \eta$ 
with $i\eta$ being a small imaginary number, 
the imaginary part of the self-energy becomes 
\begin{eqnarray}
\Im \Sigma_{\vect k} (\varepsilon) \approx - \frac{\pi}{N} \sum_{\vect p} \int d \omega 
\Bigl\{ {\cal F}_0 + 2 {\cal F}_1 \bigl( k^2 + \vect k \cdot \vect p \bigr) \Bigr\}^2
B_{\vect p} (\omega)
\bigl\{b(\omega) + f(\omega + \varepsilon) \bigr\}
\delta(\varepsilon+\omega-\varepsilon_{\vect k + \vect p}),
\end{eqnarray}
with $B_{\vect p}(\omega)= \frac{1}{\pi} \frac{\omega/\Gamma}{\bigl\{ \delta + A |\vect p - \vect Q|^2 \bigr\}^2 +(\omega/\Gamma)^2}$. 

As in the SH conductivity, 
we focus on the low-energy part $\varepsilon=0$ and momenta that satisfy the nesting condition, i.e., $\vect k$ and $\vect k + \vect Q$ are on the Fermi surface.  
Changing the momentum variable from $\vect p - \vect Q$ to $\vect q$ and 
approximate $\varepsilon_{\vect k + \vect p} =\varepsilon_{\vect k + \vect q + \vect Q} \approx \hbar {\vect v}_{\rm F} \cdot \vect q$, 
one finds 
\begin{eqnarray}
\Im \Sigma_{\vect k} (0) \approx - \frac{\pi}{N} \sum_{\vect q} 
\Bigl[ F_0 + 2 F_1 \bigl\{ k^2 + \vect k \cdot (\vect q + \vect Q) \bigr\} \Bigr]^2
B_{\vect q} (\hbar {\vect v}_{\rm F} \cdot \vect q)
\bigl\{b(\hbar {\vect v}_{\rm F} \cdot \vect q) + f(\hbar {\vect v}_{\rm F} \cdot \vect q) \bigr\}. 
\label{eq:tauAFM}
\end{eqnarray}

Neglecting the small contribution from $\vect k \cdot \vect q$, one obtains for $\vect k$, 
which is on the Fermi surface and satisfies the nesting condition, 
\begin{eqnarray}
\Im \Sigma_{\vect k} (0)  \approx -  \bigl\{ F_0 + 2 F_1 ( k_{\rm F}^2 + \vect k \cdot \vect Q)\bigr\}^2  I_{\rm AFM} (T,\delta) = - \hbar / 2 \tau_{\rm sf}. 
\end{eqnarray}
With the parabolic band $\varepsilon_{\vect k} = \frac{\hbar^2 k^2}{2m}- \varepsilon_{\rm F}$, 
$\vect k \cdot \vect Q=const.$, therefore $\Im \Sigma_{\vect k} (0)$ is independent of $\vect k$, 
which is on the Fermi surface and satisfies the nesting condition. 

\subsubsection{FM spin fluctuation}
Similarly, the carrier lifetime due to the FM fluctuation is evaluated from the electron self-energy, 
whose diagramatic representation is given in Supplementary Figure~\ref{fig:selfenergy}. 
As discussed in Ref.~\cite{SOkamoto2019}, the derivation of the inverse carrier lifetime is identical to that by the AFM spin fluctuation 
except that $B_{\vect p}$ in Eq.~(\ref{eq:tauAFM}) is replaced by 
$B_{\vect p}(\omega)= \frac{1}{\pi} \frac{\omega/\Gamma_{\vect p}}{( \delta + A p^2 )^2 +(\omega/\Gamma_{\vect p})^2}$, which 
leads to
\begin{eqnarray}
1 / \tau_{\rm sf} = - \frac{2}{\hbar} \Im \Sigma_{\vect k} (0)  \approx  \frac{2}{\hbar} \bigl\{ F_0 + 2 F_1 k_{\rm F}^2 \bigr\}^2  I_{\rm FM} (T,\delta). 
\end{eqnarray}
In contrast to the AFM case, this lifetime is independent of momentum as long as it is on the Femi surface. 

\subsection{$\sigma_{\rm SH}$ in the quantum critical regime III in the clean limit}

Here, we briefly discuss $\sigma_{\rm SH}$ in the quantum critical regime III with no disorder effects. 
The main difference appears in the FM case, where the absence of the disorder effects implies no cutoff momentum $q_{\rm c}$ in the SF propagator. 
This leads to the modified $T$ dependence of $\delta$ and, as a result, $I_{\rm FM} (T,\delta)$ as discussed in Supplementary Note~\ref{subsection:FM}. 

When the carrier lifetime is solely from the electron-electron scattering, 
$\sigma_{\rm AFM,SH}^{\rm side \, jump}$, $\sigma_{\rm FM,SH}^{\rm side \, jump}$, and $\sigma_{\rm FM,SH}^{\rm skew \, scatt.}$ diverge as $T \rightarrow 0$ 
because the divergence of $\tau_{\vect k} \propto T^{-2}$ is stronger than that of $\tilde I_{\rm AFM} (T,\delta)$ and $I_{\rm FM} (T,\delta)$. 
When the SF contributes to the carrier lifetime, such divergence is cutoff as shown in Supplementary Figure~\ref{fig:tdepIII}. 

$\sigma_{\rm AFM,SH}^{\rm skew \, scatt.}$ in the $T$ regime III does not show a diverging behavior even with $r_{\rm dis}=0$. 
Therefore, the change in $\sigma_{\rm AFM,SH}^{\rm skew \, scatt.}$ by nonzero $r_{\rm sf}$ is small. 

\begin{figure}
\begin{center}
\includegraphics[width=0.8\columnwidth, clip]{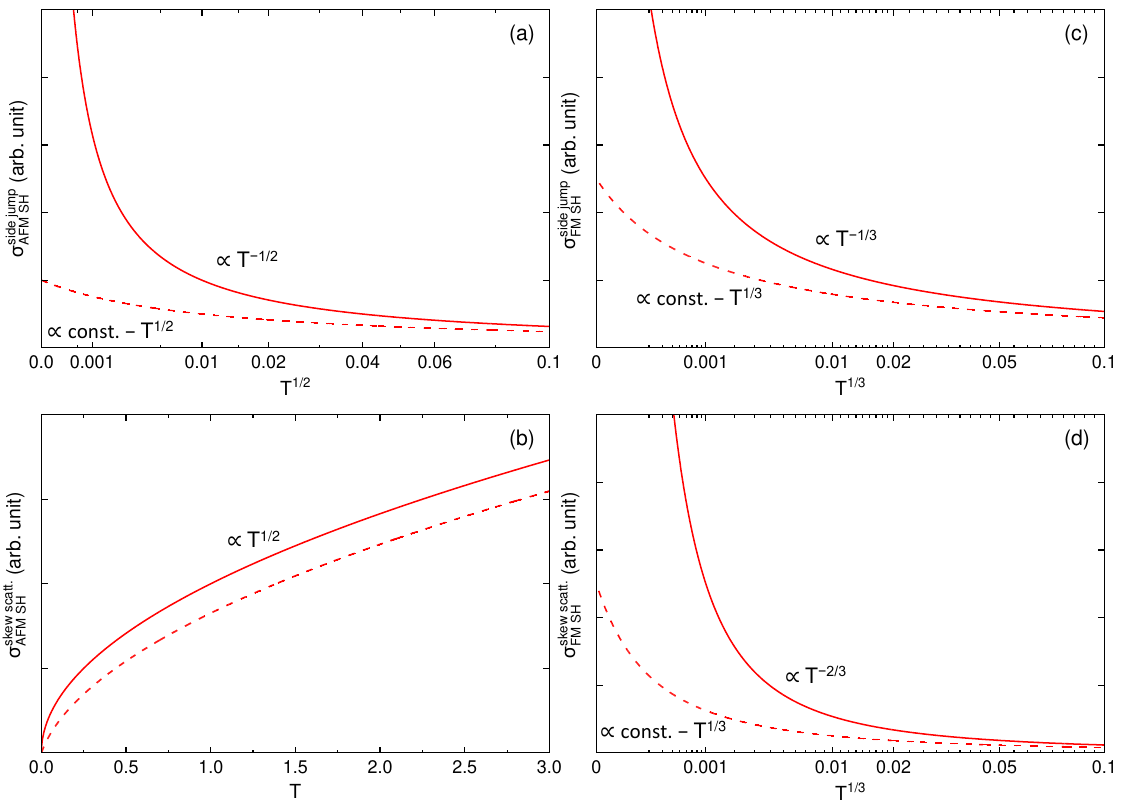}
\caption{Schematic temperature dependence of the SH conductivity with no disorder effects in the $T$ regime III with $T_{\rm N,C} \rightarrow 0$. 
(a) The side-jump type contribution $\sigma_{\rm AFM,SH}^{\rm side \, jump}$, 
(b) the skew-scattering type contribution $\sigma_{\rm AFM,SH}^{\rm skew \, scatt.}$, 
(c) the side-jump type contribution $\sigma_{\rm FM,SH}^{\rm side \, jump}$, and 
(d) the skew-scattering type contribution $\sigma_{\rm FM,SH}^{\rm skew \, scatt.}$. 
The carrier lifetime is modeled as $\tau_{\vect k}^{-1} =r_{\rm ee} T^2 + r_{\rm sf} T^{3/2}$ in (a,b), and 
$\tau_{\vect k}^{-1} =r_{\rm ee} T^2 + r_{\rm sf} T^{5/3}$ in (c,d). 
For the spin-fluctuation contribution, $r_{\rm sf}=0$ is used for solid lines and $\tau_{\rm sf}=0.1$ for dashed lines.
}
\label{fig:tdepIII}
\end{center}
\end{figure}


\begin{thebibliography}{*}

\bibitem{Dyakonov1971a}D’yakonov, M. I.  \& Perel, V. I. Possibility of orienting electron spins with current. JETP Lett. {\bf 13}, 467-469  (1971). 
\bibitem{Dyakonov1971b}D’yakonov, M. I.  \& Perel, V. I. Current-induced spin orientation of electrons in semiconductors. Phys. Lett. 35A, 459-460 (1971).

\bibitem{Hirsch1999} Hirsch, J. E. Spin Hall Effect. Phys. Rev. Lett. {\bf 83}, 1834-1837 (1999). 

\bibitem{Saitoh2006}Saitoh, E. Ueda, M. Miyajima, H. \& Tatara, G. Conversion of spin current into charge current at room temperature: Inverse spin-Hall effect. Appl. Phys. Lett. {\bf 88}, 182509 (2006). 

\bibitem{Maekawa2006}Maekawa, S. Concepts in Spin Electronics (Oxford University Press,Oxford, U.K., 2006).

\bibitem{Murakami2011}Murakami, S. \& Nagaosa, N. Spin Hall Effect. 
Comprehensive Semiconductor Science and Technology 1,  Elsevier, 222-278 (2011)

\bibitem{Sinova2015}Sinova, J. Valenzuela, S. O. Wunderlich, J. Back, C. H. \& Jungwirth, T. Spin Hall effects. Rev. Mod. Phys. {\bf 87}, 1213-1259 (2015). 

\bibitem{Nagaosa2010}Nagaosa, N. Sinova, J. Onoda, S. MacDonald, A. H. \& Ong, N. P. Anomalous Hall effect. Rev. Mod. Phys. {\bf 82}, 1539-1592 (2010). 

\bibitem{Sinova2004}Sinova, J. Culcer, D. Niu, Q. Sinitsyn, N. A. Jungwirth, T. \& MacDonald, A. H. Universal Intrinsic Spin Hall Effect. Phys. Rev. Lett. {\bf 92}, 126603 (2004). 

\bibitem{Murakami2004}Murakami, S. Nagaosa, N. \& Zhang, S.-C. Spin-Hall Insulator. Phys. Rev. Lett. {\bf 93}, 156804 (2004). 

\bibitem{Smit1955}Smit, J. The spontaneous hall effect in ferromagnetics I. Physica (Amsterdam) {\bf 21}, 877-887 (1955).
\bibitem{Smit1958}Smit, J. The spontaneous hall effect in ferromagnetics II. Physica (Amsterdam) {\bf 24}, 39-51 (1958). 

\bibitem{Berger1970}Berger, L. Side-Jump Mechanism for the Hall Effect of Ferromagnets. Phys. Rev. B {\bf 2}, 4559-4566 (1970). 
\bibitem{Berger1972}Berger, L. Application of the Side-Jump Model to the Hall Effect and Nernst Effect in Ferromagnets. Phys. Rev. B {\bf 5}, 1862-1870 (1972). 

\bibitem{Crepieux2001}Cr{\'e}pieux, A. \& Bruno, P. Theory of the anomalous Hall effect from the Kubo formula and the Dirac equation. Phys. Rev. B {\bf 64}, 014416 (2001). 

\bibitem{Tse2006}Tse, W.-K. \& Das Sarma, S. Spin Hall Effect in Doped Semiconductor Structures. Phys. Rev. Lett. {\bf 96}, 056601 (2006). 

\bibitem{Wang2016}Wang, L. Wesselink, R. J. H. Liu, Y. Yuan, Z. Xia, K. \& Kelly, P. J. Giant Room Temperature Interface Spin Hall and Inverse Spin Hall Effects. Phys. Rev. Lett. {\bf 116}, 196602 (2016). 

\bibitem{Gorini2015}Gorini, C. Eckern, U. \& Raimondi, R. Spin Hall Effects Due to Phonon Skew Scattering. Phys. Rev. Lett. {\bf 115}, 076602 (2015). 

\bibitem{Karnad2018}
Karnad, G. V. et al. Evidence for phonon skew scattering in the spin Hall effect of platinum. 
Phys. Rev. B {\bf 97}, 100405(R) (2018). 

\bibitem{Xiao2019}Xiao, C. Liu, Y. Yuan, Z. Yang, S. A. \& Niu, Q. Temperature dependence of side-jump spin Hall conductivity. Phys. Rev. B {\bf 100}, 085425 (2019). 

\bibitem{Hoffmann2013}Hoffmann, A. Spin Hall Effects in Metals. IEEE Trans. Magn.{\bf 49}, 5172-5193 (2013).

\bibitem{Okamoto2019}Okamoto, S. Egami, T. \& Nagaosa, N. Critical Spin Fluctuation Mechanism for the Spin Hall Effect. Phys. Rev. Lett. {\bf 123}, 196603 (2019). 

\bibitem{Moriya1985}Moriya, T. Spin Fluctuations in Itinerant Electron Magnetism, Solid-State Sciences {\bf 56} (Springer-Verlag, Berlin, 1985). 

\bibitem{Moriya2000}Moriya, T. \& Ueda, K. Spin fluctuations and high temperature superconductivity. Adv. Phys. {\bf 49}, 555-606 (2000). 

\bibitem{Ueda1975}Ueda, K. \& Moriya, T. Contribution of Spin Fluctuations to the Electrical and Thermal Resistivities of Weakly and Nearly Ferromagnetic Metals. J. Phys. Soc. Jpn. {\bf 39}, 605-615 (1975). 

\bibitem{Ueda1977}Ueda, K. Electrical Resistivity of Antiferromagnetic Metals. J. Phys. Soc. Jpn. {\bf 43}, 1497-1508 (1975). 

\bibitem{Wei2012}Wei, D.H. Niimi, Y. Gu, B. Ziman, T. Maekawa, S. \& Otani, Y. The spin Hall effect as a probe of nonlinear spin fluctuations. Nat. Commun. {\bf 3}, 1058 (2012). 

\bibitem{Ou2018}Ou, Y. Ralph, D. C. \& Buhrman, R. A.  
Strong Enhancement of the Spin Hall Effect by Spin Fluctuations near the Curie Point of Fe$_x$Pt$_{1-x}$ Alloys. 
Phys. Rev. Lett. {\bf 120}, 097203 (2018). 

\bibitem{Wu2022}Wu, P.-H. Qu, D. Tu, Y.-C. Lin, Y.-Z. Chien, C. L. \& Huang, S.-Y. Exploiting Spin Fluctuations for Enhanced Pure Spin Current. Phys. Rev. Lett. {\bf 128}, 227203 (2022). 

\bibitem{Gu2012}Gu, B. Ziman, T. \& Maekawa, S. Theory of the spin Hall effect, and its inverse, in a ferromagnetic metal near the Curie temperature. Phys. Rev. B {\bf 86}, 241303(R) (2012). 

\bibitem{Kondo1962}Kondo, J. Anomalous Hall Effect and Magnetoresistance of Ferromagnetic Metals. Prog. Theor. Phys. {\bf 27}, 772-792 (1962). 

\bibitem{Du2014}Du, C. Wang, H. Yang, F. \& Hammel, P. C. Systematic variation of spin-orbit coupling with $d$-orbital filling: Large inverse spin Hall effect in $3d$ transition metals. Phys. Rev. B {\bf 90}, 140407(R) (2014). 

\bibitem{Qu2015}Qu, D. Huang, S. Y. \& Chien, C. L. Inverse spin Hall effect in Cr: Independence of antiferromagnetic ordering. Phys. Rev. B {\bf 92}, 020418(R) (2015). 

\bibitem{Fang2023}
Fang, C. et al. Observation of the Fluctuation Spin Hall Effect in a Low-Resistivity Antiferromagnet. 
Nano Lett. {\bf 23}, 11485 (2023).


\bibitem{Anderson1961}Anderson, P. W. Localized Magnetic States in Metals. Phys. Rev. {\bf 124}, 41-53 (1961).

\bibitem{Kondo1964}Kondo, J. Resistance Minimum in Dilute Magnetic Alloys. Prog. Theor. Phys. {\bf 32}, 37-49 (1964). 

\bibitem{Moriya1973a}Moriya, T. \& Kawabata, A. Effect of Spin Fluctuations on Itinerant Electron Ferromagnetism. J. Phys. Soc. Jpn. {\bf 34}, 639-651 (1973). 
\bibitem{Moriya1973b}Moriya, T. \& Kawabata, A.Effect of Spin Fluctuations on Itinerant Electron Ferromagnetism. II.  J. Phys. Soc. Jpn. {\bf 35}, 669-676 (1973). 

\bibitem{Hertz1976}Hertz, J. A. Quantum critical phenomena. Phys. Rev. B {\bf 14}, 1165-1184 (1976). 

\bibitem{Millis1993}Millis, A. J. Effect of a nonzero temperature on quantum critical points in itinerant fermion systems. Phys. Rev. B {\bf 48}, 7183-7196 (1993). 

\bibitem{Nagaosa1999}Nagaosa, N. Quantum Field Theory in Strongly Correlated Electron Systems (Springer-Verlag, Berlin, 1999). 

\bibitem{Schrieffer1989}Schrieffer, J. R. Wen, X. G. \& Zhang, S. C. Dynamic spin fluctuations and the bag mechanism of high-$T_c$ superconductivity. Phys. Rev. B {\bf 39}, 11663-11679 (1989). 

\bibitem{Woelfle2021}W{\"o}lfle, P. \& Ziman, T. Theory of record thermopower near a finite temperature magnetic phase transition in IrMn. Phys. Rev. B {\bf 104}, 054441 (2021). 

\bibitem{Baber1937}Baber, W. G. The contribution to the electrical resistance of metals from collisions between electrons. Proc. R. Soc. A {\bf 158}, 383-396 (1937). 

\bibitem{Bloch1930}Bloch, F. Z. To the electrical resistance law at low temperatures. Physik {\bf 59}, 208-214 (1930). 

\bibitem{Ziman1960}Ziman, J.M. Electrons and Phonons: The Theory of Transport Phenomena in Solids (Oxford, 1960). 

\bibitem{Lohneysen1996}L{\"o}hneysen, H. v. Non-Fermi-liquid behaviour in the heavy-fermion system CeCu$_{6-x}$Au$_x$. J. Phys. Condens. Matter {\bf 8}, 9689-9706 (1996).

\bibitem{Lohneysen2007}L{\"o}hneysen, H. v. Rosch, A. Vojta, M. \& W{\"o}lfle, P. Fermi-liquid instabilities at magnetic quantum phase transitions. Rev. Mod. Phys. {\bf 79}, 1015-1075 (2007).

\bibitem{Balz2018}Baltz, V. Manchon, A. Tsoi, M. Moriyama, T. Ono, T. \& Tserkovnyak, Y. Antiferromagnetic spintronics. Rev. Mod. Phys. {\bf 90}, 015005 (2018). 

\bibitem{Kato2022}
Kato, Y. D. et al. Optical anomalous Hall effect enhanced by flat bands in ferromagnetic van der Waals semimetal. 
npj Quantum Mater. {\bf 7}, 73 (2022). 

\end{thebibliography}

\begin{thebibliography}{*}

\bibitem[1]{SKondo1962}Kondo, J. Anomalous Hall Effect and Magnetoresistance of Ferromagnetic Metals.  Prog. Theor. Phys. {\bf 27}, 772-792 (1962). 

\bibitem[2]{SOkamoto2019}Okamoto, S. Egami, T. \& Nagaosa, N. Critical Spin Fluctuation Mechanism for the Spin Hall Effect. Phys. Rev. Lett. {\bf 123}, 196603 (2019). 

\bibitem[3]{SMoriya1985}Moriya, T. Spin Fluctuations in Itinerant Electron Magnetism, Solid-State Sciences {\bf 56} (Springer-Verlag, Berlin, 1985). 

\bibitem[4]{SUeda1975}Ueda, K. \& Moriya, T. J. Phys. Soc. Jpn. {\bf 39}, 605-615 (1975). 

\bibitem[5]{SNagaosa1999}Nagaosa, N. Quantum Field Theory in Strongly Correlated Electron Systems (Springer-Verlag, Berlin, 1999).

\bibitem[6] {SWoelfle2021}W{\"o}lfle, P. \& Ziman, T. Phys. Rev. B {\bf 104}, 054441 (2021). 

\bibitem[7]{SUeda1977}Ueda, K. J. Phys. Soc. Jpn. {\bf 43}, 1497-1508 (1997). 

\end{thebibliography}
\end{document}